%% file: STDM-2017-18-PAPER.tex
\newcommand*{\ATLASLATEXPATH}{latex/}
\pdfoutput=1
\documentclass[cernpreprint, texlive=2016, UKenglish]{\ATLASLATEXPATH atlasdoc}
\usepackage[biblatex=true,subfigure]{\ATLASLATEXPATH atlaspackage}
\usepackage{\ATLASLATEXPATH atlasbiblatex}
\usepackage{\ATLASLATEXPATH atlasphysics}
\graphicspath{{logos/}{figures/}}

\usepackage{diagbox}

\input{STDM-2017-18-PAPER-metadata}

\hypersetup{pdftitle={ATLAS document},pdfauthor={The ATLAS Collaboration}}

\begin{document}

\maketitle

\tableofcontents

\section{Introduction}
\label{sec:intro}
\input{introduction}

\section{ATLAS detector and data samples}
\label{sec:atlasdet_MC}
\input{detector}
\input{samples}

\section{Selection of $Z(\nu\bar{\nu})\gamma$ events}
\label{sec:selection}
\input{selection}
\section{Background estimation}
\label{sec:backgrounds}
\input{bkg_est}

\section{Integrated and differential cross sections}
\label{sec:crossection}
\input{cross_sec}

\section{Limits on triple gauge-boson couplings}
\label{sec:atgc}
\input{atgc}

\FloatBarrier
\section{Conclusion}
\label{sec:summary}

\input{summary}

\section*{Acknowledgements}

\input{acknowledgements/Acknowledgements}

\clearpage
\printbibliography

\clearpage
\input{atlas_authlist}

\end{document}

%% file: STDM-2017-18-PAPER-metadata.tex
\AtlasTitle{Measurement of the $Z\gamma\to\nu\bar{\nu}\gamma$ production cross section in $pp$ collisions at $\sqrt{s} =13~\TeV$ with the ATLAS detector and limits on anomalous triple gauge-boson couplings}

\AtlasAbstract{
The production of $Z$ bosons in association with a high-energy photon ($Z\gamma$ production) is studied in the neutrino decay channel of the $Z$ boson using $pp$ collisions at $\sqrt{s}$~=~13~\TeV. The analysis uses a data sample with an integrated luminosity of 36.1~fb$^{-1}$ collected by the ATLAS detector at the LHC in 2015 and 2016. Candidate $Z\gamma$ events with invisible decays of the $Z$ boson are selected by requiring significant transverse momentum ($\pT$) of the dineutrino system in conjunction with a single isolated photon with large transverse energy ($\ET$). The rate of $Z\gamma$ production is measured as a function of photon $\ET$, dineutrino system $\pT$ and jet multiplicity. Evidence of anomalous triple gauge-boson couplings is sought in $Z\gamma$ production with photon $\ET$ greater than 600~\GeV. No excess is observed relative to the Standard Model expectation, and upper limits are set on the strength of $ZZ\gamma$ and $Z\gamma\gamma$ couplings.
}

\AtlasRefCode{STDM-2017-18}

\PreprintIdNumber{CERN-EP-2018-220}

\AtlasJournalRef{JHEP 12 (2018) 010}

\AtlasDOI{10.1007/JHEP12(2018)010}

\AtlasCoverSupportingNote{ATL-COM-PHYS-2018-248}{https://cds.cern.ch/record/2309495}

\AtlasCoverCommentsDeadline{16 September 2018}

\AtlasCoverEgroupEditors{atlas-stdm-2017-18-editors@cern.ch}

\AtlasCoverEgroupEdBoard{atlas-stdm-2017-18-editorial-board@cern.ch}

%% file: introduction.tex
The production of a $\Zboson$ boson in association with a photon in proton--proton ($pp$) collisions has been studied at the Large Hadron Collider (LHC) since the beginning of 
its operation in 2010~\cite{STDM-2011-17,STDM-2012-07,CMS-SMP-13-014,CMS-SMP-14-019,STDM-2014-01}. These studies have been used to test the electroweak 
sector of the Standard Model (SM) and to search for new physics effects, such as potential couplings of $\Zboson$ bosons to photons. Previous publications from 
experiments at LEP~\cite{Achard:2004ds,Abdallah:2007ae,Abbiendi:2000cu,Abbiendi:2004bf,Schael:2009zz} and the Tevatron~\cite{Abazov:2011qp,Abazov:2009cj,Aaltonen:2011zc} have shown no evidence for anomalous properties of neutral gauge bosons at the LHC. 
The set of data from the second period of the LHC operation provides the opportunity for more accurate measurements of the diboson production rate in $pp$ collisions, and facilitates higher-precision tests of triple gauge-boson couplings (TGCs). 

This paper presents a measurement of $\Zboson\gamma$ production with the $Z$ boson decaying into neutrinos. The analysis uses 36.1~fb$^{-1}$ of $pp$ collision data collected 
with the ATLAS detector\footnote{ATLAS uses a right-handed coordinate system with its origin at the nominal interaction point (IP) in the centre of the detector and the $z$-axis along the beam pipe. The $x$-axis points from the IP to the centre of the LHC ring, and the $y$-axis points upward. Cylindrical coordinates $(r,\phi)$ are used in the transverse plane, $\phi$ being the azimuthal angle around the $z$-axis. The pseudorapidity is defined in terms of the polar angle $\theta$ as $\eta=-\ln\tan(\theta/2)$. The angular distance is measured in units of $\Delta R \equiv \sqrt{(\Delta\eta)^{2} + (\Delta\phi)^{2}}$.} 
at the LHC, operating at a centre-of-mass energy of 13~\TeV. 
The measurements are made both with no restriction on the system recoiling against the $\Zboson\gamma$ pair (inclusive events) and by requiring that no jets with $|\eta|<4.5$ and $\pT$~>~50~\GeV~(exclusive events) are present in addition to the $\Zboson\gamma$ pair.

\begin{figure}[hbtp]
\begin{center}
\centering
\mbox{
\begin{subfigure}[]{}
\includegraphics[scale=0.3,angle=0]{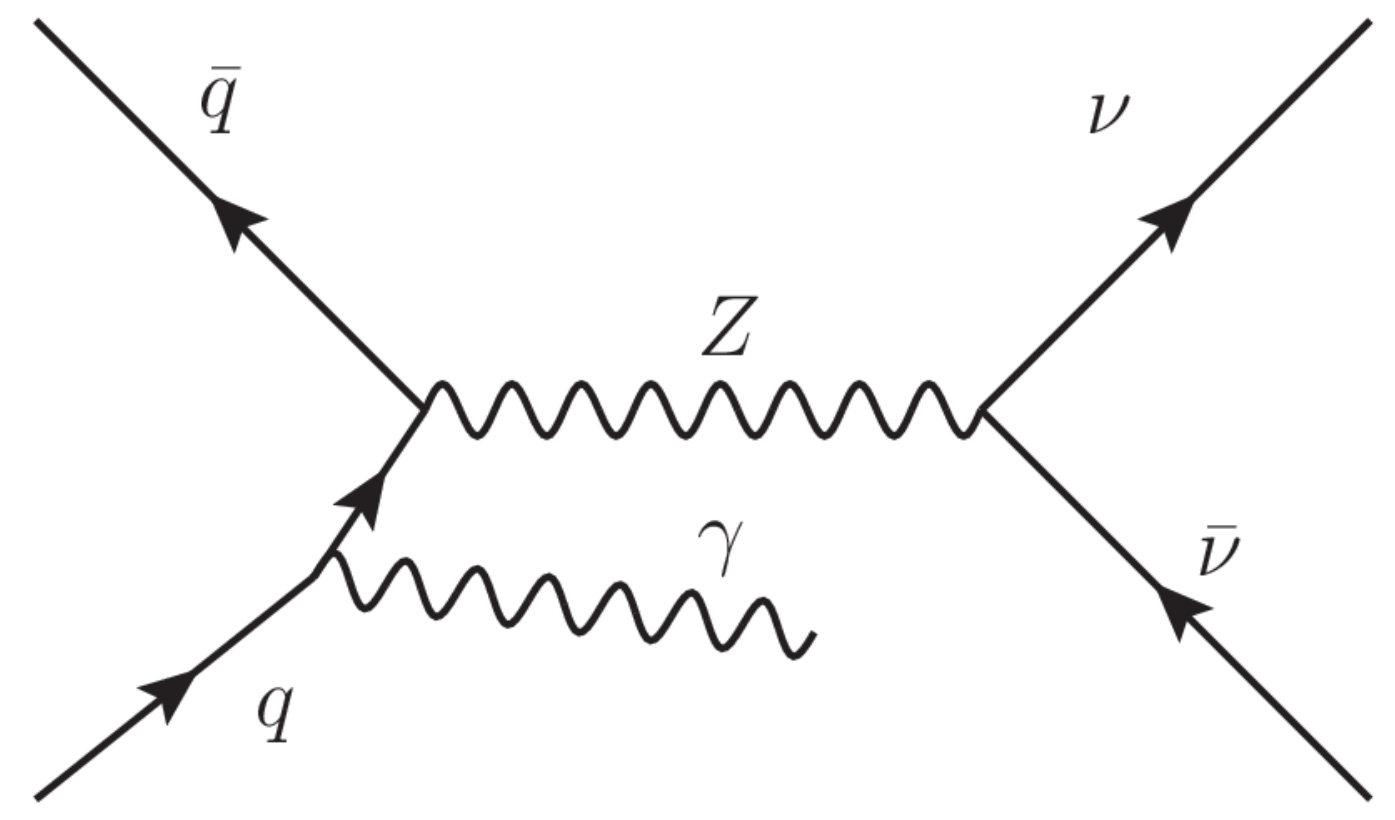}
\end{subfigure}
\begin{subfigure}[]{}
\includegraphics[scale=0.3,angle=0]{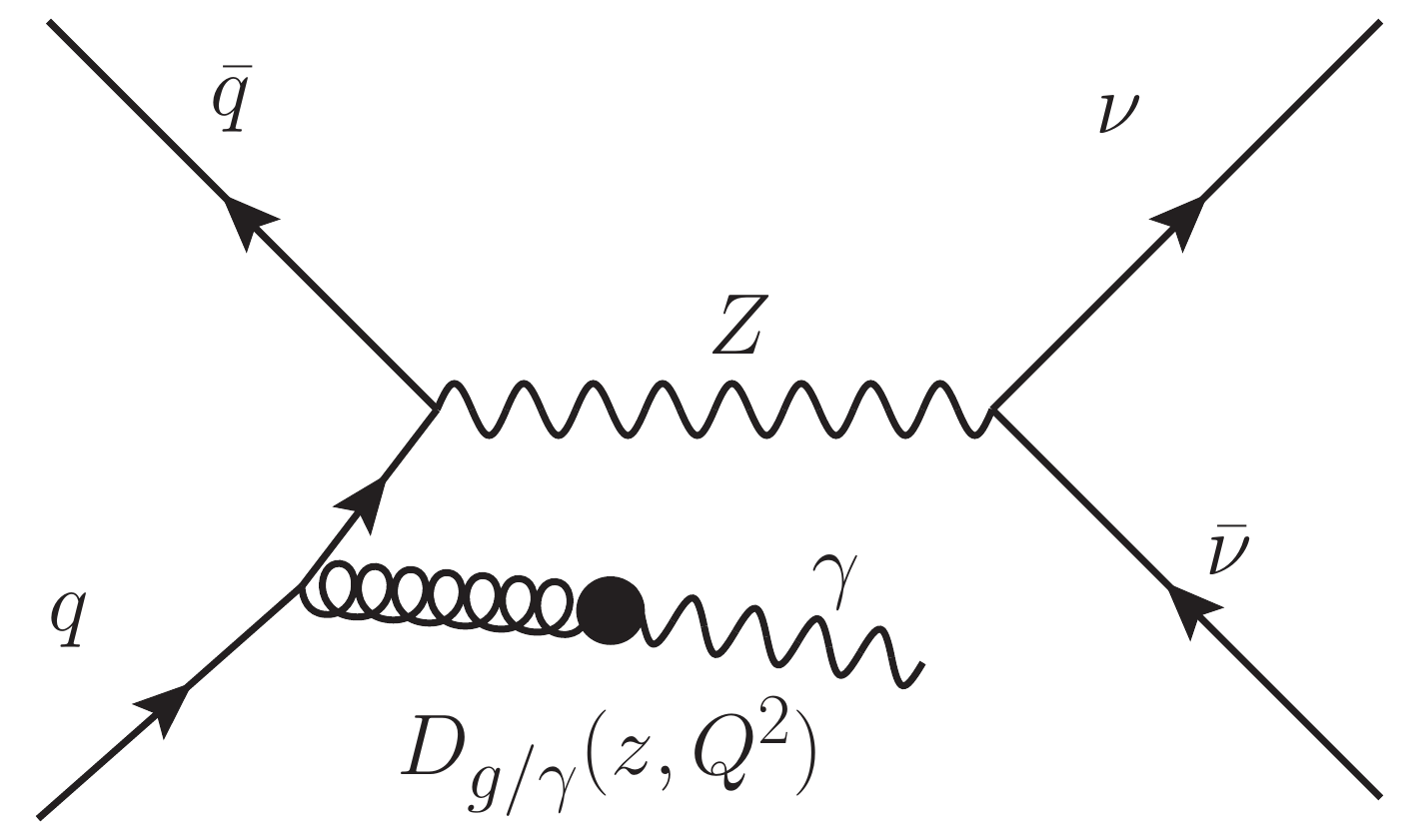}
\end{subfigure}
}
\mbox{
\begin{subfigure}[]{}
\includegraphics[scale=0.3,angle=0]{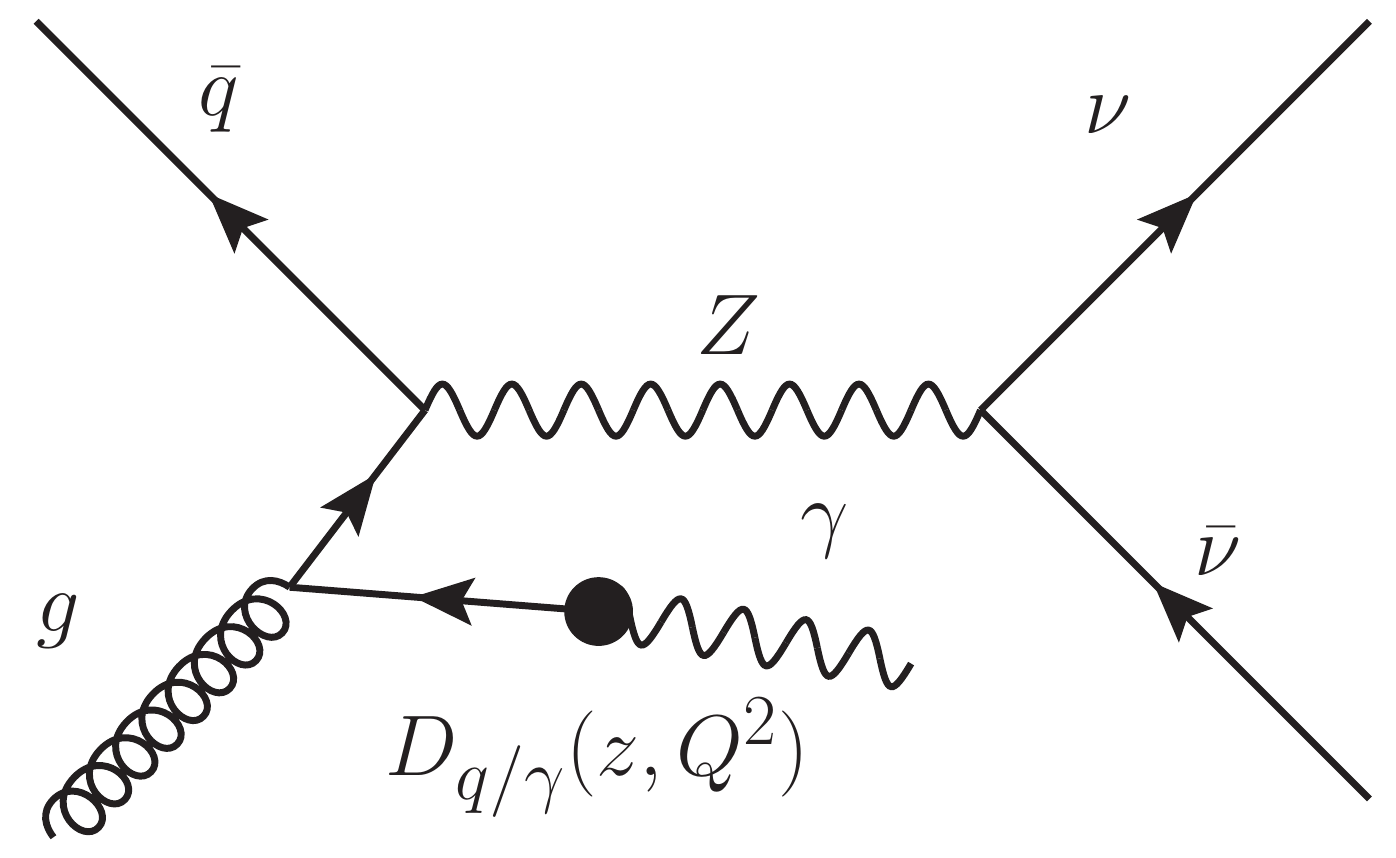}
\end{subfigure}
\begin{subfigure}[]{}
\includegraphics[scale=0.3,angle=0]{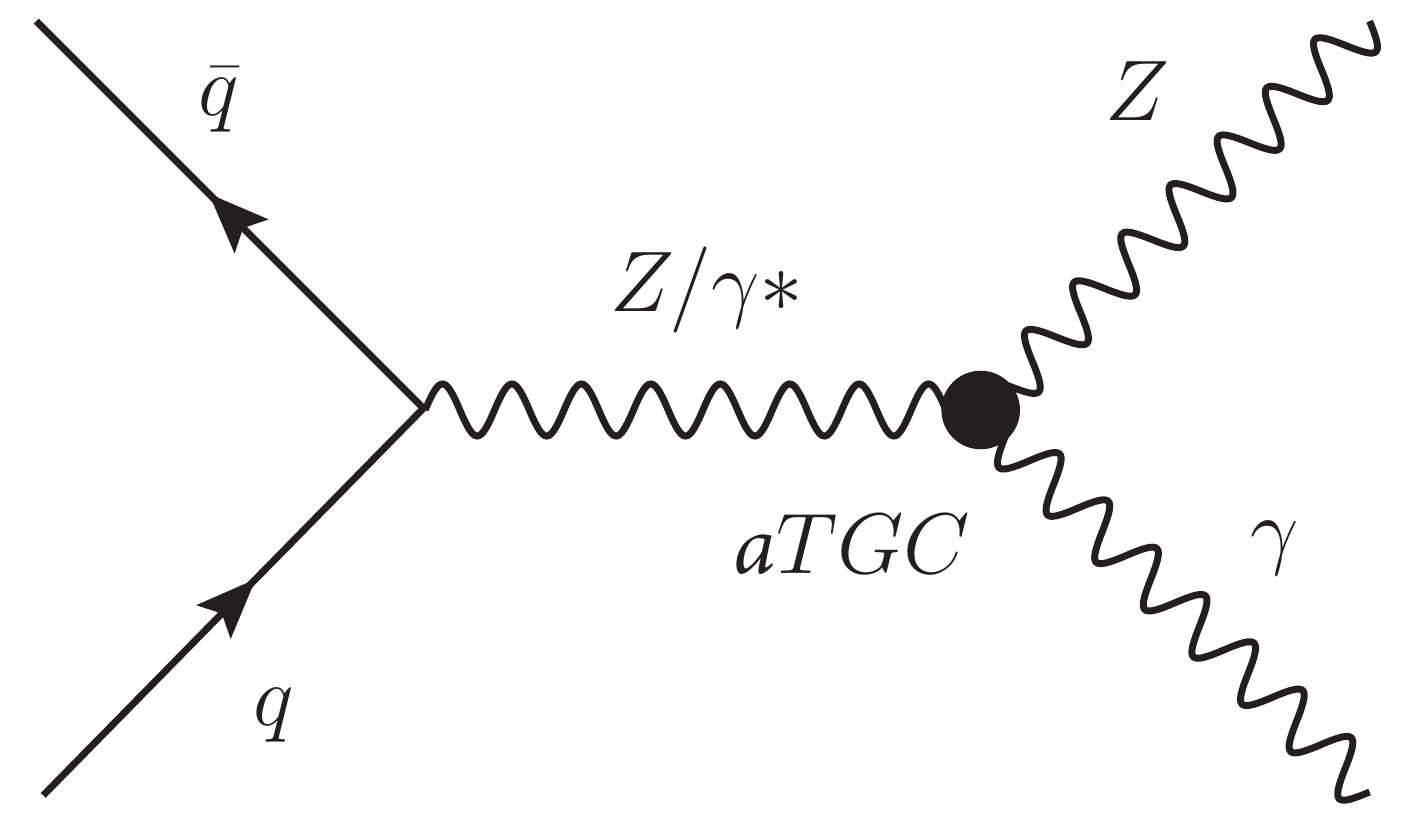}
\end{subfigure}
}
\caption{Feynman diagrams of $Z(\nu\bar{\nu})\gamma$ production: (a) initial-state photon radiation (ISR); (b,c) contributions from the $Z+q(g)$ processes in which a photon emerges from the fragmentation of a quark or a gluon; and (d) an aTGC vertex.}
\label{fig:feynmanZgg}
\end{center}
\end{figure}

The $\nu\bar{\nu}\gamma$ final state in the SM can be produced by a $Z$ boson decaying into neutrinos in association with photon emission from initial-state quarks or from quark/gluon fragmentation. 
These processes are illustrated by the leading-order Feynman diagrams shown in Figures~\ref{fig:feynmanZgg}(a)--(c). 
An example of an anomalous triple gauge-boson coupling (aTGC) of $\Zboson$ bosons 
and photons is shown in Figure~\ref{fig:feynmanZgg}(d). Such couplings are forbidden at tree level in the SM but can arise in theories that extend the SM~\cite{bib:atgc_technicolor,bib:atgc_mssm}.

A study of the $Z(\nu\bar{\nu})\gamma$ process has several advantages over processes with $Z$ decay into hadrons or charged leptons.
The channel with hadrons in the final state is contaminated by a large multijet background. A higher $Z$ boson branching ratio into neutrinos
relative to that into charged leptons provides an opportunity to study the $Z\gamma$ production in a more energetic (higher $E^{\gamma}_\mathrm{T}$) region, where the sensitivity of this process to bosonic couplings is higher~\cite{STDM-2014-01,Aaboud:2017pds}.
In addition, the neutrino channel is sensitive to anomalous neutrino dipole moments, although a higher integrated luminosity than that available to this study would be required to
significantly improve upon LEP results~\cite{Gutierrez-Rodriguez:2017bzv,Acciarri:1997im}.

The measurements of the rate and kinematic properties of the $Z\gamma$ production from this study 
are compared with SM predictions obtained from two higher-order perturbative parton-level calculations at next-to-leading order (NLO) and 
next-to-next-to-leading order (NNLO) in the strong coupling constant $\alphas$, as well as with a parton shower Monte Carlo (MC) simulation. 
The measured $Z\gamma$ production cross section at high values of photon $\ET$ is used to search for aTGCs 
($ZZ\gamma$ and $Z\gamma\gamma$). For these searches an exclusive selection is used, providing higher sensitivity to the anomalous couplings due 
to further background suppression.

%% file: detector.tex
\subsection{ATLAS detector and experimental data set}
\label{sec:detector}

The ATLAS detector at the LHC is described in detail in Ref.~\cite{PERF-2007-01}. A short overview is presented here, with an emphasis 
on the subdetectors needed for a precision measurement of the $Z(\nu\bar{\nu})\gamma$ final state. The ATLAS detector covers nearly the entire solid angle surrounding the collision point. Its major components are an
inner tracking detector (ID) surrounded by a thin superconducting solenoid providing a 2~T axial magnetic field,
electromagnetic (ECAL) and hadron (HCAL) calorimeters, and a muon spectrometer (MS). 
The ID is composed of three subsystems. Two detectors cover the pseudorapidity range $|\eta|<2.5$: the silicon pixel detector and the silicon microstrip tracker (SCT). The outermost system of the ID, with an acceptance of $|\eta|<2.0$, 
is composed of a transition radiation tracker (TRT). The TRT provides identification information for electrons by the detection of transition radiation. 
The MS is composed of three large superconducting air-core toroid magnets, a system of three stations of chambers for tracking
measurements, with high precision in the range $|\eta|<2.7$, and a muon trigger system covering the range $|\eta|<$ 2.4.

The ECAL is composed of alternating layers of passive lead absorber interspersed with active liquid-argon gaps.
It covers the range of |$\eta$| < 3.2 and plays a crucial role in photon identification. 
For |$\eta$| $<$ 2.5 the calorimeter has three longitudinal layers in shower depth, with the first layer having the highest
granularity in the $\eta$ coordinate, and the second layer collecting most of the electromagnetic shower energy for high-$p_{\mathrm{T}}$ objects.  
A thin presampler layer precedes the ECAL over the range $|\eta|<1.8$, and is used to correct for the energy lost by EM particles upstream of the calorimeter.
The HCAL, surrounding the ECAL, is based on two different technologies, 
with scintillator tiles or liquid-argon as the active medium, and with either steel, copper, or tungsten as the absorber material. 
Photons are identified as narrow, isolated showers in the ECAL with no penetration into the HCAL.
The fine segmentation of the ATLAS calorimeter system allows an efficient separation of jets from isolated prompt photons.

Collision events are selected using a hardware-based first-level trigger and a software-based high-level trigger. The resulting
recorded event rate from LHC $pp$ collisions at $\sqrt{s}$~=~13~\TeV\ during the data-taking period in 2015 and 2016 was approximately 1~kHz \cite{TRIG-2016-01-corr}.
After applying criteria to ensure good ATLAS detector operation, the total integrated luminosity useful for data analysis is 36.1~fb$^{-1}$.
The uncertainty in the combined 2015+2016 integrated luminosity is 2.1$\%$. It is derived, following a methodology similar to that detailed in Ref.~\cite{DAPR-2013-01}, and using the LUCID-2 detector for the baseline luminosity measurements \cite{LUCID2}, from calibration of the luminosity scale using $x$--$y$ beam-separation scans.

%% file: samples.tex
\subsection{Simulation of signal and backgrounds}
\label{sec:samples}

Simulated signal and background events were produced with various Monte Carlo event generators, processed through a full ATLAS detector simulation~\cite{SOFT-2010-01} 
using \textsc{Geant4}~\cite{bib-geant4}, and then reconstructed with the same procedure used for data.
Additional $pp$ interactions (pile-up), in the same and neighbouring bunch crossings, were overlaid on the hard-scattering process in the MC simulation.
The MC events were then reweighted to reproduce the distribution of the number of interactions per bunch crossing observed in data.

For the signal modeling $\SHERPA$~2.2.2~\cite{SherpaREF} with the NNPDF3.0 NNLO PDF set~\cite{Ball:2014uwa} is used as the baseline event generator.
The signal sample was generated with up to three additional final-state partons at leading order (LO) and up to one additional final-state parton at next-to-leading order (NLO).
Alternative signal samples, the first generated using $\SHERPA$~2.1.1 with the CT10 PDF set~\cite{Lai:2010vv} and the second generated using {\textsc{MG5\_aMC@NLO}}~2.3.3~\cite{madgraph} with the NNPDF3.0 NLO PDF set and interfaced to the $\PYTHIA$~8.212~\cite{Sjostrand:2007gs} parton shower model, are considered for studies of systematic uncertainties. Signal samples with non-zero anomalous triple gauge-boson couplings were also generated using $\SHERPA$~2.1.1 with the CT10 PDF set. The values of coupling constants used in the generation are chosen to be equal to the expected limits obtained in a previous ATLAS study~\cite{STDM-2014-01}.

Background events containing $Z$ bosons with associated jets were simulated using $\SHERPA$~2.1.1 with the CT10 PDF set, while background events containing $W$ bosons with associated jets were simulated using $\SHERPA$~2.2.0 with the NNPDF3.0 NNLO PDF set. For both of these processes the matrix elements were calculated for up to two partons at NLO and four partons at LO. Background events containing a photon with associated jets were simulated using $\SHERPA$~2.1.1 with the CT10 PDF set. Matrix elements were calculated with up to four partons at LO. Background events containing a lepton pair and a photon with associated jets were simulated using $\SHERPA$~2.2.2 with the NNPDF3.0 NNLO PDF set. Matrix elements including all diagrams with three electroweak couplings were calculated for up to one parton at NLO and up to three partons at LO.

%% file: selection.tex
The event selection criteria are chosen to provide precise cross-section measurements of $Z(\nu\bar{\nu})\gamma$ production and good sensitivity to anomalous gauge-boson couplings between photons and $Z$ bosons. The selection is optimized for obtaining a high signal efficiency together with good background rejection. 

Events are required to have been recorded with stable beam conditions and with all relevant detector subsystems operational. Event candidates in both data and MC simulation are selected using the lowest-$\ET$ unprescaled single-photon trigger: this requires the presence of at least one cluster of energy deposition in the ECAL with transverse energy $\ET$ larger than 140~\GeV, satisfying
the \textit{loose} identification criteria described in Ref.~\cite{ATL-PHYS-PUB-2016-014}.
The trigger efficiency is greater than 98\% for photons selected for this analysis. 

\subsection{Object selection}
\label{sec_objsel}
Photon candidates are reconstructed~\cite{PERF-2013-04} from ECAL energy clusters with $|\eta|<2.37$ and $\ET>150$~\GeV. They are classified either as converted (candidates with a matching reconstructed conversion vertex or a matching track consistent with having originated from a photon conversion) or as unconverted (all other candidates). Both kinds of photon candidates are used in the analysis. Electron candidates are reconstructed~\cite{ATLAS-CONF-2016-024} from ECAL energy clusters with $|\eta|<2.47$ that are associated with a reconstructed track in the ID with transverse momentum $\pT>7$~\GeV. The ECAL cluster of the electron/photon candidate must lie outside the transition region between the barrel and endcap (\mbox{$1.37<|\eta|<1.52$}). Muon candidates are reconstructed from tracks in the MS that have been matched to a corresponding track in the inner detector, and are referred to as ``combined muons''. The combined track is required to have $\pT>7$~\GeV\ and $|\eta|<2.7$.

The shower shapes produced in the ECAL are used to identify photons and electrons. Photons are required to pass all the requirements on shower shape variables which correspond to the \textit{tight} photon identification criteria~\cite{ATL-PHYS-PUB-2016-014}. The \textit{tight} photon identification efficiency ranges from 88\% (96\%) to 92\% (98\%) for unconverted (converted) photons with $\pT>100$~\GeV. A sample of ``preselected'' photons, used for the calculation of missing transverse momentum, are required to satisfy the less restrictive \textit{loose} identification criteria of Ref.~\cite{ATL-PHYS-PUB-2016-014}. Electron candidates are required to satisfy \textit{loose}~\cite{ATLAS-CONF-2016-024} electron identification criteria, whose efficiency is greater than 84\%. Muon candidates are required to satisfy \textit{tight} identification criteria as described in Ref.~\cite{PERF-2015-10}, with efficiency greater than 90\% for combined muons used in the selection.

Electron and muon candidates are required to originate from the primary vertex\footnote{Each primary vertex candidate is reconstructed from at least two associated tracks with $\pT>0.4$~\GeV. The primary vertex is selected among the primary vertex candidates as the one with the highest sum of the squared transverse momenta of its associated tracks.} by demanding that the significance of the transverse impact parameter, defined as the absolute value of the track’s transverse impact parameter, $d_0$, measured relative to the beam trajectory, divided by its uncertainty, $\sigma_{d_{0}}$, satisfy $|d_{0}|/\sigma_{d_{0}} < 3$ for muons and $|d_{0}|/\sigma_{d_{0}} < 5$ for electrons. The difference $z_0$ between the value of the $z$ coordinate of the point on the track at which $d_0$ is defined, and the longitudinal position of the primary vertex, is required to satisfy $|z_0\cdot \mathrm{sin}(\theta)| < 0.5$~mm for both the muons and electrons.

Photon, electron and muon candidates are required to be isolated from other particles. The following criteria are used for photons: the total transverse energy in ECAL energy clusters within $\Delta R = 0.4$ of the photon candidate is required to be less than 2.45~\GeV~+~$0.022\cdot\ET^{\gamma}$, and the scalar sum of the transverse momenta of the tracks located within a distance $\Delta R = 0.2$ of the photon candidate is required to be less than $0.05\cdot\pT^{\gamma}$. 
For preselected photons, isolation criteria are not applied. For muons and electrons, the isolation requirement is based on track information and is tuned to have an efficiency of at least 99\%~\cite{PERF-2015-10}.

Jets are reconstructed from topological clusters in the calorimeter~\cite{PERF-2014-07} using the anti-$k_t$ algorithm~\cite{Cacciari:2008gp} with a radius parameter of $R = 0.4.$ Events with jets arising from detector noise or other non-collision sources are discarded~\cite{ATLAS-CONF-2015-029}. A multivariate combination of track-based variables is used to suppress jets originating from pile-up in the ID acceptance~\cite{PERF-2014-03}. 
The energy of each jet is calibrated and corrected for detector effects using a combination of simulated events and in situ methods~\cite{PERF-2016-04} using data collected at $\sqrt{s} = 13$~\TeV. The selected jets are required to have $\pT$ larger than 50~\GeV\ and $|\eta|<4.5$. 

The missing transverse momentum is defined as the negative vector sum of the transverse momenta of all reconstructed physics objects in the event~\cite{Aaboud:2018tkc} (leptons with $\pT>7$~\GeV, preselected photons with $\pT>10$~\GeV\ and jets with $\pT>20$~\GeV), plus a ``soft term'' incorporating tracks from the primary vertex that are not associated with any such objects~\cite{ATL-PHYS-PUB-2015-027}. The resulting vector is denoted $\vec{E}_{\mathrm{T}}^{\mathrm{miss}}$ since it includes calorimetric energy measurements, and its magnitude $\met$ is used as a measure of the total transverse momentum of neutrinos in the event.

To resolve ambiguities in the object reconstruction, jet candidates lying within $\Delta R=0.3$ of the photon candidates are removed.

\subsection{Signal region definition}
The signal region (SR) is defined to have exactly one \textit{tight} isolated photon, as described above. In order to reduce the contamination from events that do not contain high-energy neutrinos (mainly $\gamma$~+ jet background with fake $\met$ from jet momenta mismeasurements) the selected events are required to have $\met>150$~\GeV. To reduce the number of $W(\ell\nu)\gamma$ and $Z(\ell\ell)\gamma$ events, a lepton veto is applied: events with any selected electrons or muons are discarded. 
A requirement of at least $10.5$~\GeV$^{1/2}$ for the $\met$ significance, defined as $\met{}/\sqrt{\Sigma \pT^{\mathrm{jet}}+\ET^{\gamma}}$, further suppresses background contributions with fake $\met$. An additional angular separation requirement $\Delta\phi(\vec{E}_{\mathrm{T}}^{\mathrm{miss}},\gamma)>\pi/2$ is made, which suppresses the $pp\rightarrow W(e\nu)+X$ background. These object and event selection requirements define the reconstruction-level fiducial region and are summarized in Table~\ref{table:ZgSigReg}. 

\begin{table}[hbtp]
	\begin{center}
		\begin{tabular}{ccc}
			\hline
			Photons\hspace{1cm}  & Leptons\hspace{1cm} & Jets\\
			\hline
			$\ET > 150$~\GeV\hspace{1cm} & $\pT>7$~\GeV\hspace{1cm} & $\pT > 50$~\GeV\\
			$|\eta| <  2.37$,\hspace{1cm} & $|\eta|<2.47(2.7)$ for $e(\mu)$,\hspace{1cm} & $|\eta| < $ 4.5\\
			excluding $1.37<|\eta|<1.52$\hspace{1cm} & excluding $1.37<|\eta^e|<1.52$\hspace{1cm} & $\Delta R(\mathrm{jet},\gamma) > 0.3$\\
			\hline
			\multicolumn{3}{c}{Event selection}\\ \hline
			\multicolumn{3}{c}{$N^\gamma=1$,\hspace{0.2cm} $N^{e,\mu}=0$,\hspace{0.2cm} $\met>150$~\GeV,\hspace{0.2cm} $\met\ \mathrm{signif.}>10.5$~\GeV$^{1/2}$,\hspace{0.2cm} $\Delta\phi(\vec{E}_{\mathrm{T}}^{\mathrm{miss}},\gamma)>\pi/2$}\\
			\multicolumn{3}{c}{Inclusive : $N_{\mathrm{jet}} \geq 0$,\hspace{0.2cm} Exclusive : $N_{\mathrm{jet}} = 0$}\\ \hline
		\end{tabular}
		\caption{Definition of the fiducial region. The object selection is presented in the top part of the table, while the event selection is described in the bottom part.}
		\label{table:ZgSigReg}
	\end{center}
\end{table}

To simplify the interpretation of the results and comparison with theory predictions, the cross section is measured in an extended fiducial region, defined at particle level\footnote{``Particle level'' quantities are defined in terms of stable particles in the MC event record with a proper decay length $c\tau > 10$~mm which are produced from the hard scattering, including those that are the products of hadronization. The particle-level jets are reconstructed using the anti-$k_t$ algorithm with a radius parameter of $R=0.4$, using all stable particles except for muons and neutrinos. The particle-level jets in ATLAS do not include muons because jets are built from calorimeter clusters, excluding muons.}
in Table~\ref{table:ZgRegions_extfid}. Compared with the fiducial region, the extended fiducial region removes requirements on $\met$ significance, $\Delta\phi(\vec{E}_{\mathrm{T}}^{\mathrm{miss}},\gamma)$, the lepton veto and the transition $\eta$ region for photons.
In the signal event selection at particle level, the $\met$ significance and $\Delta\phi(\vec{E}_{\mathrm{T}}^{\mathrm{miss}},\gamma)$ are given by $p^{\nu\bar{\nu}}_\mathrm{T}{}/\sqrt{\Sigma \pT^{\mathrm{jet}}+\ET^{\gamma}}$ 
and $\Delta\phi(\vec{p}^{\;\nu\bar{\nu}}_\mathrm{T},\gamma)$, respectively. Photon isolation at the particle level is performed using the same requirements and cone sizes as described for the reconstruction-level isolation in Section~\ref{sec_objsel}.

\begin{table}[hbtp]
        \begin{center}
        \begin{tabular}{lc}
        \hline
        Category  & Requirement \\
        \hline
        Photons & $\ET^{\gamma}$ $>$ 150~\GeV \\
                & $|\eta| < $ 2.37 \\
        \hline
        Jets & $|\eta| < $ 4.5 \\
             & $\pT$ $>$ 50~\GeV \\
             & $\Delta R(\mathrm{jet},\gamma) >$ 0.3 \\
             & Inclusive : $N_{\mathrm{jet}} \geq 0$, Exclusive : $N_{\mathrm{jet}} = 0$\\
        \hline
        Neutrino & $\pT^{\nu\bar{\nu}}>150$~\GeV \\
        \hline
        \end{tabular}
        \caption{Definition of the extended fiducial region. At particle level, $\pT^{\nu\bar{\nu}}$ is the equivalent of $\met$.}
        \label{table:ZgRegions_extfid}
        \end{center}
        \end{table}

%% file: bkg_est.tex
\label{sec:bkgnngnngg}

Backgrounds to the $Z(\nnbar)\gamma$ signal originate from several sources. The dominant sources (listed in decreasing order of importance) are estimated with data-driven techniques: electroweak processes such as $W(\ell\nu)\gamma$, where the lepton is not detected; events with prompt photons and mismeasured jet momenta that gives rise to missing transverse momentum; events with real $\met$ from neutrinos (such as $\Zboson(\nnbar)$ or $W(e\nu)$) and misidentified photons from either electrons or jets. The procedures used to estimate these backgrounds closely follow those of the previous ATLAS measurement~\cite{STDM-2014-01}. 
A less important source is $\ell\ell\gamma$ (mainly $\tau\tau\gamma$) production, which is estimated from MC simulation and is expected to contribute roughly 1\% of the selected event yield. 
In the following, each source of background is discussed in detail together with the method used for its estimation.

Misidentified events from $W(\ell\nu)\gamma$ production are one of the dominant background contributions. 
A large fraction (about 60\%) of this contamination arises from $W(\tau\nu)\gamma$ events. 
Photon+jets events form another sizeable background contribution to the signal region.
For the estimation of these backgrounds, two control regions (CRs) are defined by selecting events with the same criteria used for the SR but requiring either exactly one charged lepton ($e$ or $\mu$) in the event, or requiring the $\met$ significance to be less than 10.5~$\GeV^{1/2}$. The first CR is enriched with $W(\ell\nu)\gamma$ events (about 77\%) while the second CR is enriched with $\gamma$+jets events (about 55\%). The use of the 1-lepton ($e$ or $\mu$) control region for the estimation of the $W(\ell\nu)\gamma$ background to the signal region, where $\ell$ can be any of $e$, $\mu$ or $\tau$, relies on the assumption of lepton flavour universality.
A simultaneous fit to the background-enriched CRs is performed to allow the CR data to constrain the yield of these main backgrounds, initially estimated with MC simulation, by establishing the normalization factors for the $W(\ell\nu)\gamma$ and $\gamma$+jets background contribution as described in Refs.~\cite{EXOT-2015-05,STDM-2014-01}. The same background normalization factors are assumed in the CR and SR and the fit uncertainties on these factors accounts for the uncertainty from this assumption. The normalization factor for the $W\gamma$ background is found to be close to one, while the normalization factor for the $\gamma$+jets background is $1.7\pm0.5$, since the pre-fit expectation is computed at LO, for which higher-order corrections would be expected to be considerable. The pre-fit kinematic distributions of these backgrounds are taken from the MC simulation. The variations of the background yield in each bin due to each of the experimental and MC modelling uncertainties reported in Section~\ref{xsec_intro} are treated as Gaussian-distributed nuisance parameters in the likelihood function fit used to obtain the final background predictions in the SR. The dominant systematic uncertainties in the $W(\ell\nu)\gamma$ process come from MC modelling (mostly due to the QCD scale uncertainty) and from the uncertainty in the electron--photon energy scale. Their contributions are 5.8\% and 3.8\%, respectively. The systematic uncertainty for $\gamma$+jets events is also dominated by the QCD scale component, and amounts to approximately 19\%.

Misidentification of electrons as photons also contributes to the background yield in the signal region. The main source of this background is the inclusive $W(e\nu)$ process, but 
contributions also arise from the single top-quark and $t\bar{t}$ production processes.
The estimation of the size of these background contributions is done in two steps. The first is the determination of the probability for an electron to
be misidentified as a photon using $\Zboson(\ee)$ decays reconstructed as $e+\gamma$, as described in Refs.~\cite{EXOT-2015-05,STDM-2014-01}. 
The probability of observing an $e+\gamma$ pair with invariant mass near the $\Zboson$ boson mass is used to determine an electron-to-photon fake factor $f_{e\rightarrow\gamma}$.
The fake factor is found to vary between 0.6\% to 2.7\%, depending on the photon's $\eta$ and $\pT$.
The second step is the construction of a control region by applying the nominal $\nnbar\gamma$ selection criteria described in Section~\ref{sec:selection}, with the exception that an 
electron is required instead of the final-state photon, leading to a control region dominated by the $W(e\nu)$+jets process.
The estimated background is then given by the number of events in the chosen control sample scaled by the electron-to-photon fake factor.
The statistical uncertainty is determined by the size of the control sample and does not exceed 5\%. The systematic uncertainty for this background varies from 13\% to 25\%, depending on the photon $\pT$ and $\eta$, and is dominated by the difference between the fake rates obtained from $Z(ee)$ and $W(e\nu)$ MC events. This source of systematic uncertainty on the fake factor is estimated from MC simulation in order to avoid double counting the uncertainty associated with the estimation of backgrounds under the $\Zboson$ boson mass peak in collision data. The total relative systematic uncertainty of this background estimate is less than 15\%, since the main contribution comes from the most populated central pseudorapidity region and has $\pT<250$~\GeV, where the systematics on the fake factor is the smallest.

To estimate the contribution from background due to the misidentification of jets as photons, a two-dimensional sideband method is used, as described in Ref.~\cite{STDM-2014-01}. In this method the $Z(\nnbar)\gamma$ events are separated into one signal and three control regions.
Events in the signal region require the photon to satisfy the nominal photon isolation and \textit{tight} identification requirements, 
as described in Section~\ref{sec:selection}. The photon isolation and identification criteria are modified in order to build the
control regions, which are disjoint from each other and from the signal region. The modified photon identification criteria requires photons to pass a ``non-\textit{tight}'' identification but fail the \textit{tight} identification. The non-\textit{tight} selection criteria remove requirements on four out of the nine shower shape variables required for \textit{tight} photons; the variables that are removed from the list of requirements are those that are least correlated with calorimeter isolation~\cite{STDM-2010-08}. Two of the control regions are defined by modifying either the photon isolation or photon identification requirement, while for the third control region both the isolation and identification requirements are modified. The number of background 
events in the signal region can be derived from the number of observed events in the control regions according to the methodology described in Ref.~\cite{STDM-2014-01}. The statistical uncertainty of the background is established by the event yields in the four regions, while the systematic uncertainty is 29\% and is dominated by the size of changes to the background estimate arising from the variation of the control regions' definitions, which leads to changes exceeding the expected size of the statistical fluctuations. This systematic uncertainty also covers possible effects due to the correlation between the isolation and identification criteria.

\begin{figure}[h]
	\begin{center}
		\includegraphics[width=0.45\textwidth]{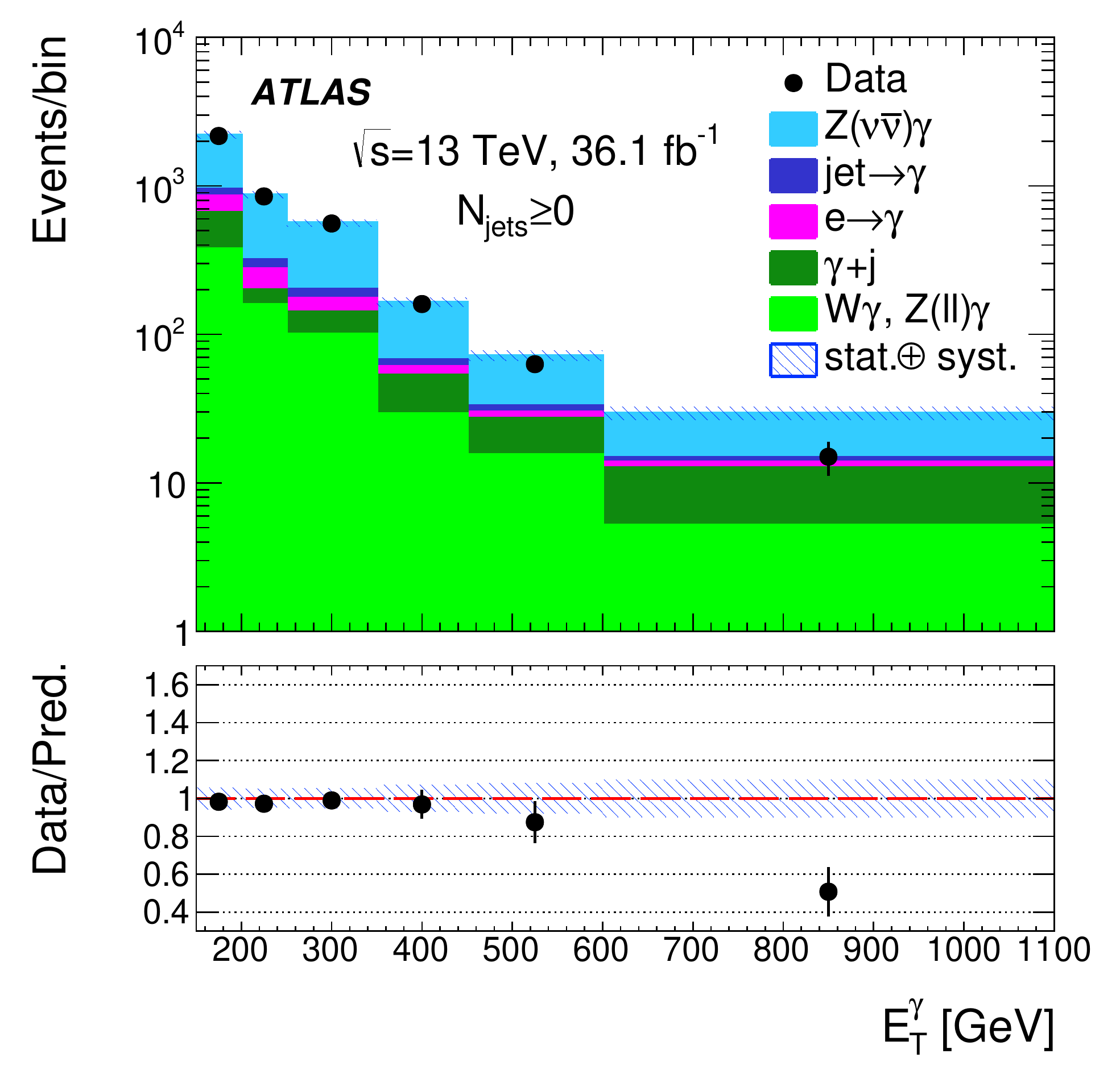}
		\includegraphics[width=0.45\textwidth]{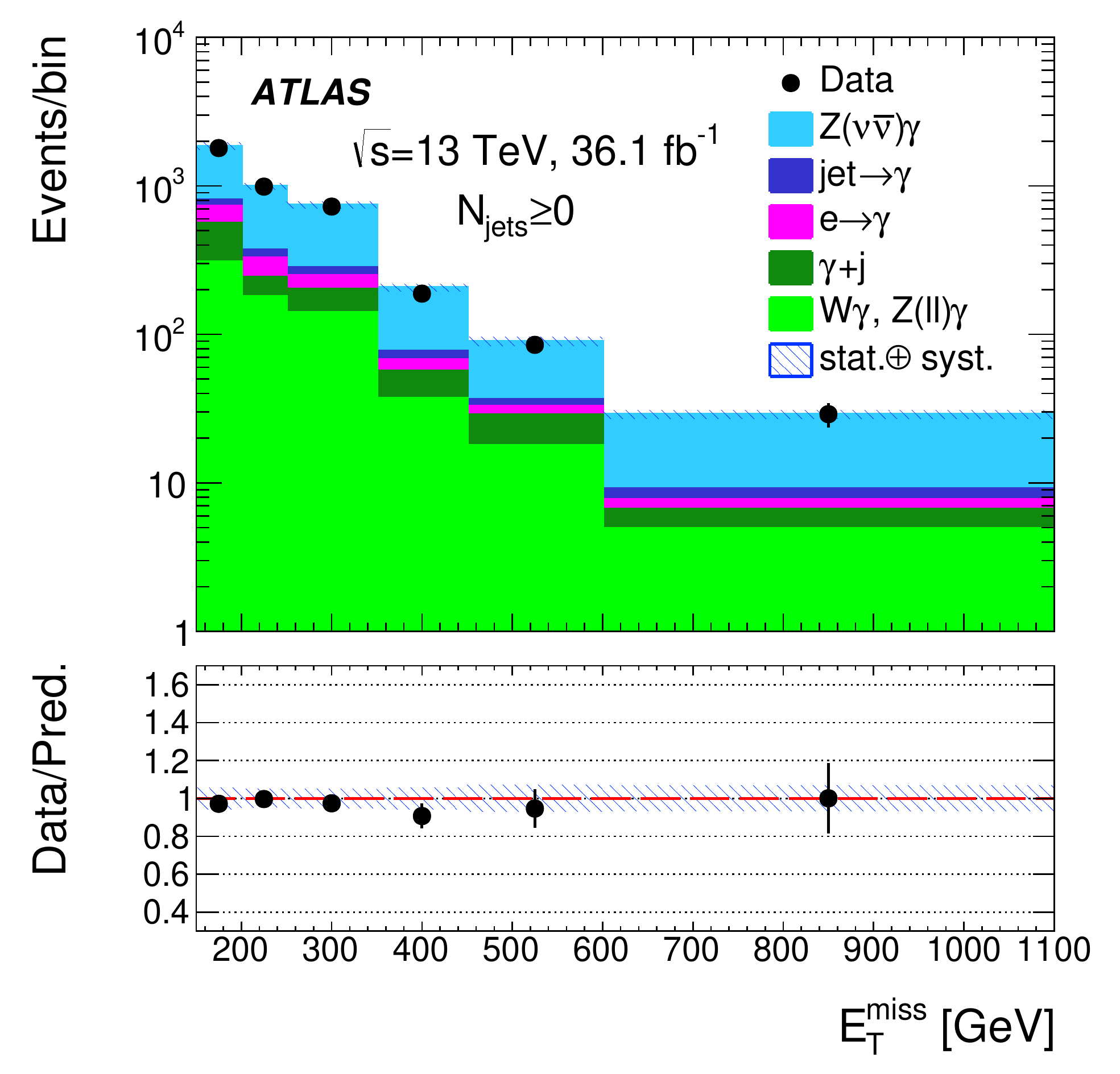}\\
		\includegraphics[width=0.45\textwidth]{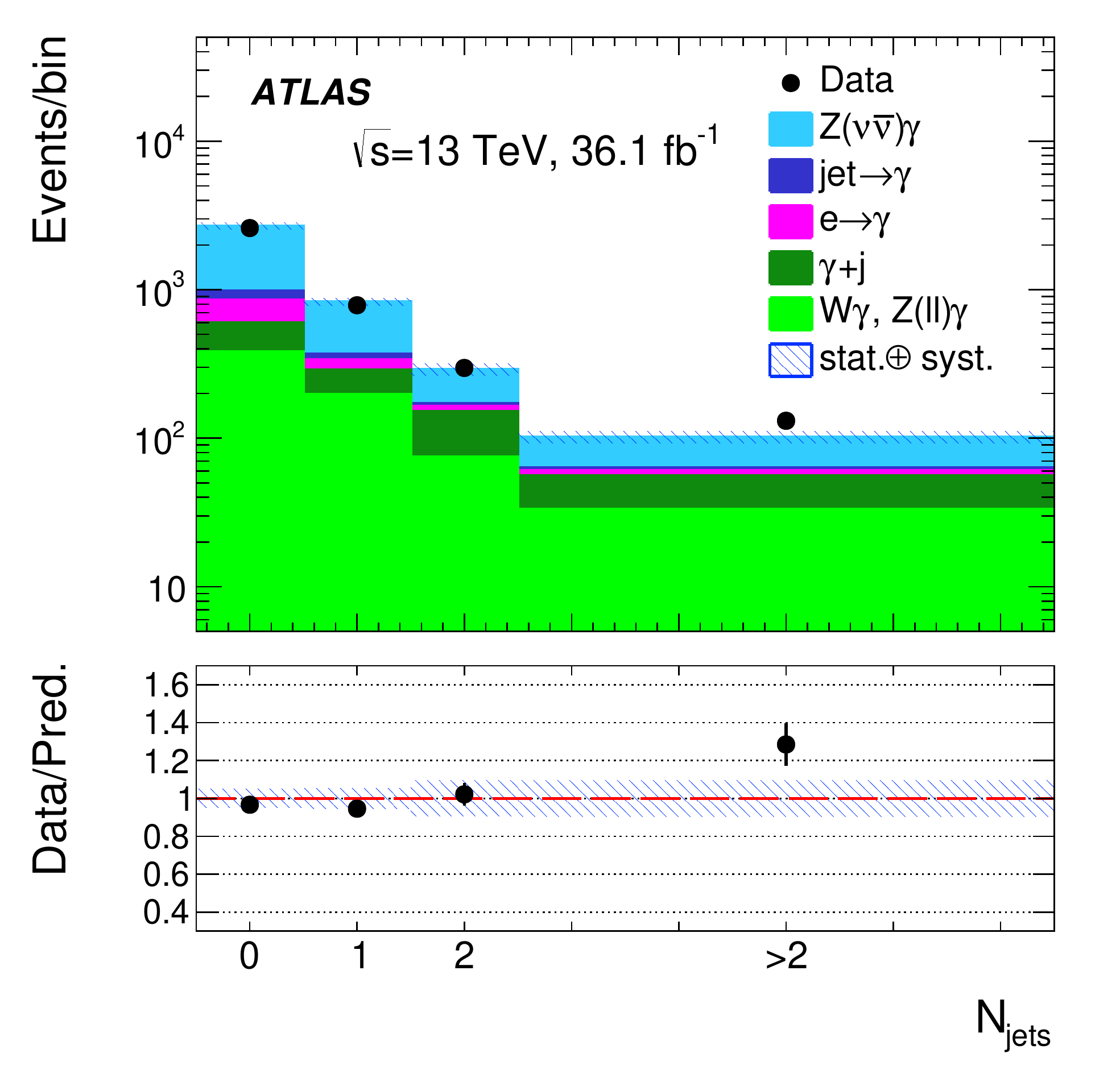}\\
\caption{Top left: photon $\ET$ distribution; top right: missing transverse momentum distribution; bottom: jet multiplicity distribution, in the inclusive ($N_{\mathrm{jets}}\geq0$) signal region. MC expectations are scaled to the integrated luminosity of the data using the expected MC cross section of each sample. The $W\gamma$ and $\gamma$+jet backgrounds are scaled by an additional normalization factor from the fit to data in the corresponding control regions. Backgrounds arising from electron or jet misidentification as a photon are estimated with the data-driven techniques described in the text. The dashed band represents the sum in quadrature of systematic and statistical uncertainties of both the background and signal expectation, and includes a contribution arising from the uncertainty in the integrated luminosity of the data sample.}
		\label{fig:SR_inclusive}
	\end{center}
\end{figure}

\begin{figure}[h]
	\begin{center}
		\includegraphics[width=0.45\textwidth]{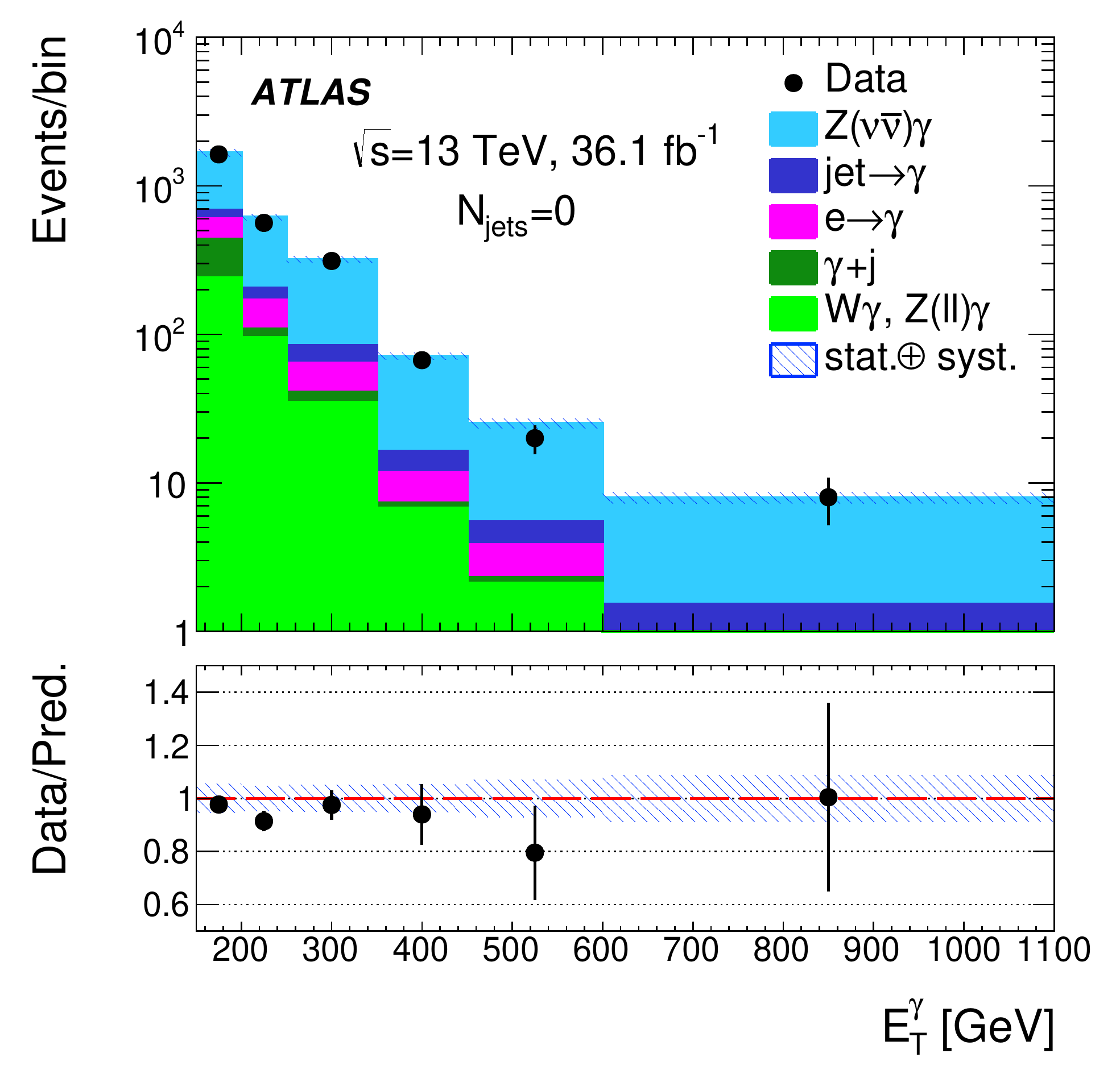}
		\includegraphics[width=0.45\textwidth]{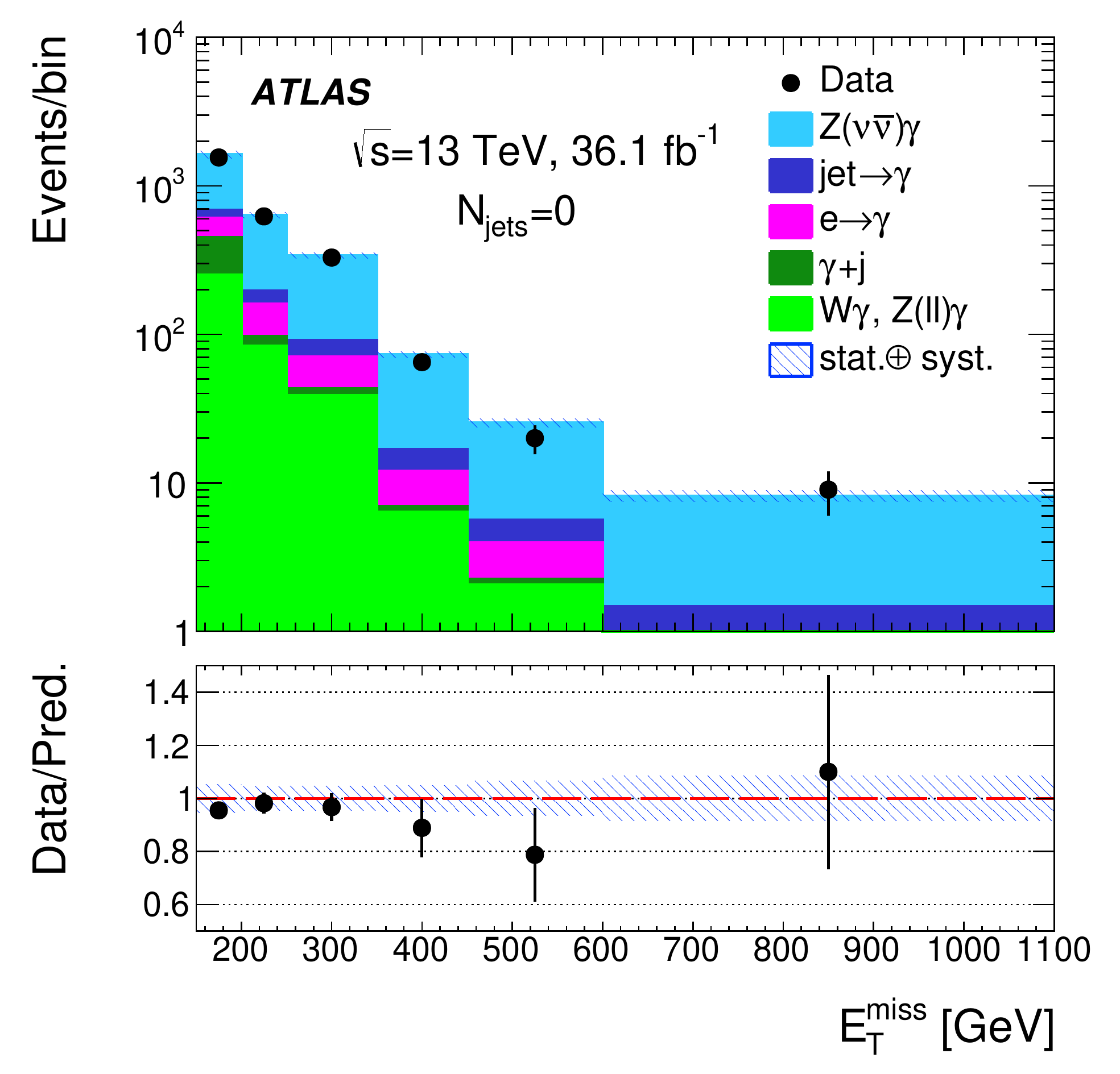}\\
\caption{Left: photon $\ET$ distribution; right: missing transverse momentum distribution, in the exclusive ($N_{\mathrm{jets}}=0$) signal region. MC expectations are scaled to the integrated luminosity of the data using the expected MC cross section of each sample. 
The $W\gamma$ and $\gamma$+jet backgrounds are scaled by an additional normalization factor from the fit to data in the corresponding control regions. Backgrounds arising from electron or jet misidentification as a photon are estimated with the data-driven techniques described in the text. The dashed band represents the sum in quadrature of systematic and statistical uncertainties of both the background and signal expectation, and includes a contribution arising from the uncertainty in the integrated luminosity of the data sample.}
                \label{fig:SR_exclusive}
	\end{center}
\end{figure}

The resulting signal and background composition is shown in Table~\ref{table:znunugresults}. 
Kinematic distributions of the photon transverse energy, missing transverse momentum, and jet multiplicity in the fiducial region for the inclusive selection ($N_{\mathrm{jets}}\geq0$) are shown in Figure~\ref{fig:SR_inclusive}. Kinematic distributions of the photon transverse energy and the missing transverse momentum in the fiducial region for the exclusive selection ($N_{\mathrm{jets}}=0$) are shown in Figure~\ref{fig:SR_exclusive}.

\begin{table}
	\begin{center}
		\begin{tabular*}{0.55\textwidth}{l@{\extracolsep{\fill}}r@{ $\pm$\extracolsep{0.01cm} }r@{ $\pm$\extracolsep{0.01cm} }r@{\extracolsep{\fill}}r@{ $\pm$\extracolsep{0.01cm} }r@{ $\pm$\extracolsep{0.01cm} }r}
			\hline
			& \multicolumn{3}{c}{N$_{\mathrm{jets}}$ $\geq$ 0} & \multicolumn{3}{c}{$N_{\mathrm{jets}}$ = 0}\\
			\hline
			$N^{W\gamma}$ &650&40&60 &360& 20& 30 \\
			$N^{\gamma+\mathrm{jet}}$ & 409&18&108 & 219& 10& 58\\
			$N^{e\rightarrow\gamma}$ & 320& 15& 45 & 254& 12& 35\\
			$N^{\mathrm{jet}\rightarrow\gamma}$ & 170&30&50& 140& 20& 40\\
			$N^{Z(\ell\ell)\gamma}$ &\extracolsep{\fill} 40& 3&3 & 26&3&2 \\
			\hline
			$N_\mathrm{total}^\mathrm{bkg}$ &1580& 50&140 & 1000& 40&90\\
			$N^\mathrm{sig} \mathrm{(exp)}$ & 2328& 4& 135 &1710&4& 91  \\
			\hline
			$N_\mathrm{total}^\mathrm{sig+bkg}$ & 3910& 50&190 & 2710& 40& 130\\
			\hline
			$N^\mathrm{data} \mathrm{(obs)}$ &\multicolumn{3}{l}{ \hspace{-3mm} 3812} &\multicolumn{3}{l}{\hspace{-3mm} 2599} \\
			\hline
		\end{tabular*}
		\caption{Summary of observed and expected yields (all backgrounds and signal) for events passing the selection requirements in data for the inclusive ($N_{\mathrm{jets}} \geq 0$) and exclusive ($N_{\mathrm{jets}}$ = 0) selections. The $W\gamma$ and $\gamma$+jet backgrounds are scaled by the normalization factor from the fit, luminosity and cross section. The $e\rightarrow\gamma$ and jet~$\rightarrow\gamma$ backgrounds are estimated using data-driven techniques. The row labelled ``$N^\mathrm{sig} \mathrm{(exp)}$'' corresponds to the $\SHERPA$ NLO prediction. The row labelled ``$N_\mathrm{total}^\mathrm{sig+bkg}$'' corresponds to the sum of the expected background contributions and expected signal. The first uncertainty is statistical, while the second is systematic. } 
		\label{table:znunugresults}
	\end{center}
\end{table}

Good agreement between data and the SM expectation is observed in the shapes of most of the measured distributions. The discrepancy in the last bin of the inclusive $\ET^{\gamma}$ distribution, which is not used to set aTGC limits, was found to be consistent with having arisen from a statistical fluctuation of the data. The uncertainties shown in the figures are treated as being uncorrelated among different systematic sources and different backgrounds.

%% file: cross_sec.tex
\label{xsec}

\subsection{Description of the cross-section measurements}
\label{xsec_intro}
The number of signal events is determined by subtracting the estimated backgrounds from the number of observed events.
The signal yield is then corrected for detection efficiencies in the fiducial region, defined in Table~\ref{table:ZgSigReg}.
The integrated cross section in the extended fiducial region, defined in Table~\ref{table:ZgRegions_extfid}, is calculated as
\begin{linenomath}
\begin{equation*}
\sigma_{\textrm{ext-fid}} = \frac{N - B}{A_{Z\gamma} \cdot C_{Z\gamma} \cdot \int{L{\textrm{d}}\,t}},
\end{equation*}
\end{linenomath}
where $N$ is the number of observed candidate events, $B$ is the expected number of background events and
$\int{L{\textrm{d}}\,t}$ is the integrated luminosity corresponding to the analyzed data set.
The factors $C_{Z\gamma}$ and $A_{Z\gamma}$ correct for detection efficiency and acceptance, respectively:
\begin{itemize}
\item $C_{Z\gamma}$ is defined as the number of reconstructed signal events satisfying all selection criteria
divided by the number of events that, at particle level, meet the acceptance criteria of the fiducial region;
\item $A_{Z\gamma}$ is defined as the number of signal events within the fiducial region
divided by the number of signal events within the extended fiducial region, with both numerator and
denominator defined at particle level.
\end{itemize}

The corrections $A_{Z\gamma}$ and $C_{Z\gamma}$ are determined using the $Z\gamma$ signal events generated by $\SHERPA$ and are summarized in Table~\ref{table:Acceptances} along with their uncertainties.

\subsection{Systematic uncertainties}
\label{sec_uncert}
Systematic uncertainties in the acceptances $A_{Z\gamma}$ are evaluated by varying the PDF sets, the value of $\alphas$, 
the renormalization and factorization scales (QCD scale uncertainty), and the Monte Carlo parameter tunes for the parton shower (PS) and multi-parton interactions (MPI).
In total, 100 error sets are checked for the NNPDF3.0 NNLO PDF variation, leading to a relative uncertainty of
$0.76\%$ for the inclusive case and $0.35\%$ for the exclusive case.
These numbers fully cover variations arising from the use of alternative PDF sets such as CT14~\cite{Dulat:2015mca} and MMHT2014~\cite{Harland-Lang:2014zoa}.
The uncertainty from $\alphas$ is estimated by varying it within the range of its world-average value as provided in Ref.~\cite{Agashe:2014kda} and is found to be negligible.
The effects of the renormalization and factorization scale uncertainties are assessed by varying these two scales independently by a factor of two from their nominal values, removing combinations where the two variations differ by a factor of four, and taking the envelope of the resulting cross-section variations as the size of the associated systematic uncertainty. Uncertainties from the PS and MPI are evaluated using a series of eigentunes for the $\PYTHIA$ generator with its A14 parameter tune~\cite{ATL-PHYS-PUB-2014-021}. The size of the uncertainty from the renormalization and factorization scales does not exceed $3.0\%$,
while PS and MPI uncertainties cause variations from $1.9\%$ to $2.7\%$ for the inclusive and exclusive cases, respectively.
The total uncertainties in the acceptance factors are summarized in Table~\ref{table:Acceptances}.

\begin{table}
\begin{center}
\begin{tabular*}{0.4\textwidth}{@{\extracolsep{\fill}}lcc}
\hline
& $N_{\mathrm{jets}} \geq 0$ & $N_{\mathrm{jets}} = 0$\\
\hline
$A_{Z\gamma}$ & $ 0.816 \pm 0.029 $  &  $0.952\pm0.026$ \\
$C_{Z\gamma}$ & $ 0.904 \pm 0.031 $  & $0.889\pm0.037$ \\
\hline

\end{tabular*}
\caption{Summary of values of the correction factors ($C_{Z\gamma}$) and acceptances ($A_{Z\gamma}$) 
for the $Z\gamma$ cross-section measurements. The uncertainty presented here includes only systematic components, since the statistical uncertainty is found to be negligible.}
\label{table:Acceptances}
\end{center}
\end{table}

Systematic uncertainties affecting the correction factor $C_{Z\gamma}$ include contributions arising from uncertainties in the efficiencies of the trigger, reconstruction and particle identification, as well as the uncertainties in the energy, momentum scales and resolutions of the final-state objects. 
Additional systematic uncertainty sources arise from the modelling of particle spectra and pile-up events. Spectrum modelling uncertainties are estimated by varying the PDF 
set and QCD scales as described above for the case of the acceptance factor $A_{Z\gamma}$. 
Some of these contributions are found to have a non-linear dependence on photon transverse energy, $\met$ or jet multiplicity. In these cases,
uncertainties estimated as a function of these observables are used in the unfolding process of Section~\ref{sec:Diff_cross_sec} when 
the corresponding kinematic distributions are derived from the signal sample.
Table~\ref{table:SystematicAcceptances} displays the size of the individual contributions 
to the uncertainties in the $C_{Z\gamma}$ factor; the total uncertainty is summarized in Table~\ref{table:Acceptances}.

\begin{table}
\begin{center}
\begin{tabular*}{0.61\textwidth}{l r@{.}l@{\extracolsep{-1.5cm}}r@{\extracolsep{0cm}.}l}
\hline
Source & \multicolumn{2}{c}{Relative uncertainty [$\%$]} \\
\hline
& \multicolumn{2}{l}{\hspace{-0.1cm}$N_{\mathrm{jets}}\geq0$} & \multicolumn{2}{c}{\hspace{-0.3cm}$N_{\mathrm{jets}}=0$}\\
\hline
Trigger efficiency				& 0&79 & 0&79 \\
Photon identification efficiency 		& 1&5 & 1&5 \\
Photon isolation efficiency      		& 0&48 & 0&47 \\
Electron--photon energy scale                    & 2&5 & 2&5  \\
Electron--photon energy resolution               & 0&11 & 0&09  \\
Jet energy scale               			& 0&92 & 2&2  \\
Jet energy resolution            		& 0&10 & 0&43  \\
$\met$ scale		                     	& $<$0&1 & $<$0&1 \\
$\met$ resolution		                	& 0&13 & $<$0&1 \\
Pile-up simulation                               & 0&85 & 1&1  \\
Spectrum modelling                 		& 1&3 & 1&3  \\
\hline
Sum						& 3&5 & 4&2  \\
\hline
\end{tabular*}
\caption{Relative systematic uncertainties in the signal correction factor $C_{Z\gamma}$ for the inclusive and exclusive $Z\gamma$ measurements.}
\label{table:SystematicAcceptances}
\end{center}
\end{table}

\subsection{Integrated extended fiducial cross section}
\label{xsec_pheno}

The measurements of the cross sections, along with their uncertainties, are based on the maximization of the profile-likelihood ratio
\begin{linenomath}
\begin{equation*}
        \Lambda(\sigma) = \frac{\mathcal{L}(\sigma, \hat{\hat{\boldsymbol{\theta}}}(\sigma))}{\mathcal{L}(\hat{\sigma}, \hat{\boldsymbol{\theta}})},
\end{equation*}
\end{linenomath}
where $\mathcal{L}$ represents the likelihood function, $\sigma$ is the cross section,
and $\boldsymbol{\theta}$ are the nuisance parameters corresponding to the sources of systematic uncertainty.
The $\hat{\sigma}$ and $\hat{\boldsymbol{\theta}}$ terms denote the unconditional maximum-likelihood estimate of the parameters, i.e., 
the parameters for which
the likelihood is maximized for both $\sigma$ and $\boldsymbol{\theta}$.
The term $\hat{\hat{\boldsymbol{\theta}}}(\sigma)$ denotes the value of $\boldsymbol{\theta}$ that maximizes $\mathcal{L}$ for a given 
value of $\sigma$.

The likelihood function is defined as
\begin{linenomath}
\begin{equation*}
        \mathcal{L}(\sigma, \boldsymbol{\theta}) =  \mathrm{Poisson}(N~|~S(\sigma, \boldsymbol{\theta}) + B(\boldsymbol{\theta})) \cdot {\mathrm{Gaussian}}( \boldsymbol{\theta_0}~|~\boldsymbol{\theta}), 
\end{equation*}
\end{linenomath}
representing the product of the Poisson probability of observing $N$ events, given expectations of $S$ for the signal 
and $B$ for the background,
multiplied by the Gaussian constraints  $\boldsymbol{\theta}$ on the systematic uncertainties, with central values $\boldsymbol{\theta_0}$ 
from auxiliary measurements, as described 
in Section~\ref{xsec_intro}.

The measured cross sections for $Z(\nu\bar{\nu})\gamma$ production in the extended fiducial region are summarized in Table~\ref{table:ZgCrossSections}, along with the theoretical predictions of the \textsc{Mcfm}~\cite{MCFM} generator described in Section~\ref{sec:SMpredictions}. The measured cross sections agree with the SM expectations to within one standard deviation. Systematic uncertainties arise from uncertainties in the acceptances and correction factors, as well as from uncertainties in the background estimates. These two sources contribute roughly equally to the uncertainty in the measured cross sections. Compared with the $Z\gamma$ measurements at $\sqrt{s} = 8$~\TeV~\cite{STDM-2014-01}, the systematic uncertainty is significantly reduced. This improvement is due primarily to the reduction of systematic uncertainty allowed by the data-driven estimate of the $\gamma$+jets and $W\gamma$ backgrounds.

An overall check of the SM predictions is done with the \textsc{Matrix} generator~\cite{Grazzini:2293309}. Cross sections obtained by \textsc{Matrix} (inclusive case: $\sigma^{\mathrm{ext.fid.}} = 78.6\pm 0.4\pm 4.4$~fb; exclusive case: $\sigma^{\mathrm{ext.fid.}} = 55.8\pm 0.3\pm 3.6$~fb, where the uncertainties are statistical and systematic, respectively) are found to be consistent with those from {\textsc{Mcfm}} to within their statistical uncertainty.

\begin{table}
\begin{center}
\begin{tabular}{cc}
\hline
$\sigma^{\mathrm{ext.fid.}}$ [fb] & $\sigma^{\mathrm{ext.fid.}}$ [fb] \\
 Measurement & NNLO \textsc{Mcfm} Prediction  \\
\hline
\multicolumn{2}{c}{$N_{\mathrm{jets}} \geq 0$ } \\
 83.7$^{+3.6} _{-3.5}$ (stat.)$ ^{+6.9} _{-6.2}$ (syst.)$ ^{+1.7} _{-2.0}$ (lumi.) & $78.1 \pm 0.2$(stat.)$\pm 4.7$(syst.) \\
\hline
\multicolumn{2}{c}{$N_{\mathrm{jets}} = 0$ }\\
  52.4$^{+2.4}_{-2.3}$ (stat.)$ ^{+4.0} _{-3.6}$ (syst.)$ ^{+1.2} _{-1.1}$ (lumi.) & $55.9 \pm 0.1$(stat.)$ \pm 3.9$(syst.) \\
\hline
\end{tabular}
\caption{Measured cross sections for $Z(\nu\bar{\nu})\gamma$ production within the extended fiducial region for a centre-of-mass energy of $\sqrt{s} = 13$~\TeV, with corresponding SM expectations obtained from the \textsc{Mcfm}~\cite{MCFM} generator at next-to-next-to-leading order in the strong coupling constant $\alphas$.
}
\label{table:ZgCrossSections}
\end{center}
\end{table}

\subsection{Standard Model calculations}
\label{sec:SMpredictions}

The resulting measurement of the rate and kinematic distributions of $\Zboson\gamma$ production is compared with SM expectations using the parton shower Monte Carlo generator $\SHERPA$ and the NNLO parton-level generators \textsc{Mcfm} and {\textsc{Matrix}}.
The NNPDF3.0 PDF set was used for the $\SHERPA$, {\textsc{Mcfm}} and {\textsc{Matrix}} generation. The values of the renormalization and factorization scales 
were set to $m_{\Zboson\gamma}$ for the {\textsc{Mcfm}} and {\textsc{Matrix}} NNLO generation of the $\Zboson\gamma$ process. 

The photon isolation criterion at the parton level is applied by considering a cone of variable opening angle $\Delta R$ (with maximum opening angle $\Delta R_{\mathrm{max}}=0.1$) centred around the photon direction, and requiring that the transverse energy flow inside that cone be always less than a given fraction of the photon $\pT$; this fraction is set to 0.1 when $\Delta R = \Delta R_{\mathrm{max}}$, and tends smoothly to zero when $\Delta R \rightarrow 0$, as described in Ref.~\cite{Frixione:1998jh}.
Due to this procedure, the contribution from photon fragmentation to the NNLO calculations of the {\textsc{Mcfm}} and {\textsc{Matrix}} SM predictions is zero. 

Events generated with $\SHERPA$, as described in Section~\ref{sec:samples}, are also compared with the particle-level measurements.
For the NNLO parton-level predictions, parton-to-particle correction factors \mbox{$C^{*(\mathrm{parton}\to\mathrm{particle})}$} must be applied in order to obtain the particle-level cross sections. 
These correction factors are computed as the ratios of the $pp\rightarrow Z\gamma$ cross sections predicted by $\SHERPA$ with hadronization and the underlying event disabled to the cross sections with them enabled.
The systematic uncertainty in the correction factors is evaluated by using a signal sample from an alternative generator ({\textsc{MG5\_aMC@NLO}}), taking the resulting change in $C^{*(\mathrm{parton}\to\mathrm{particle})}$ as the one-sided size of a symmetrized value for the uncertainty. This accounts for uncertainties in both the parton shower modelling and the description of the underlying event.
The value of $C^{*(\mathrm{parton}\to\mathrm{particle})}$ is found to be $0.87\pm0.04$ for the inclusive predictions and $0.97\pm0.04$ for the exclusive predictions. For the exclusive case, the parton-to-particle correction includes an additional contribution from the jet veto, which compensates for the difference in the photon isolation between the parton and particle levels. 
The particle-level cross sections are then obtained by multiplying the NNLO parton-level cross-section values by the $C^{*(\mathrm{parton}\to\mathrm{particle})}$ correction factors, and are displayed in Table~\ref{table:ZgCrossSections}. 

The systematic uncertainty in the expected NNLO SM cross sections arising from uncertainties in the QCD scale is estimated by varying the QCD scales by factors of 0.5 and 2.0 (separately for the renormalization and factorization scales, removing combinations where the two variations differ by a factor of four). The effect of the QCD scale uncertainty on the prediction for the first bin of the various differential cross-section measurements also accounts for uncertainties arising from the incomplete cancellation of divergences associated with soft gluon emission in fixed-order perturbative calculations of $Z\gamma$ production. This effect is appreciable because of the symmetric $E_{\mathrm{T}}^{\gamma}$ and $p_{\mathrm{T}}^{\nu\bar{\nu}}$ thresholds used in defining the SR. 
The corresponding corrections are estimated conservatively from the cited MC generators by evaluating the degree of compensation of the divergence that arises when the $p_{\mathrm{T}}^{\nu\bar{\nu}}$($E_{\mathrm{T}}^{\gamma}$) requirement is lowered to a value significantly below the value of the $E_{\mathrm{T}}^{\gamma}$($p_{\mathrm{T}}^{\nu\bar{\nu}}$) requirement of 150~\GeV. 
The systematic uncertainty due to the PDF choice is computed using the eigenvectors of the NNPDF 3.0 PDF set~\cite{Ball:2014uwa} and the envelope of the differences between the results obtained with the CT14~\cite{Dulat:2015mca} and MMHT2014~\cite{Harland-Lang:2014zoa} PDF sets, according to the PDF4LHC recommendations~\cite{Butterworth:2015oua}. \textsc{Matrix} predictions do not include the systematic uncertainty due to the PDF choice.

\subsection{Differential extended fiducial cross section}
\label{sec:Diff_cross_sec}

The measurement of the $Z\gamma$ production differential cross sections allows a comparison of experimental results with SM expectations for both the absolute rates and the shapes of kinematic distributions.
The measurements are performed as a function of several observables that are sensitive to higher-order perturbative QCD corrections~\cite{Campbell_Zg_NNLO} and to a possible manifestation of aTGCs~\cite{PhysRevD.47.4889}: photon transverse energy~($E_{\mathrm{T}}^{\gamma}$), the transverse momentum of the neutrino--antineutrino pair~($p_{\mathrm{T}}^{\nu\bar{\nu}}$), and jet multiplicity~($N_{\mathrm{jets}}$).   
The differential cross sections are defined in the extended fiducial region, and are extracted with an unfolding procedure that corrects for measurement inefficiencies and resolution effects that modify the observed distributions.
The procedure described in Ref.~\cite{STDM-2014-01} is followed, using an iterative Bayesian method~\cite{Agostini:1994zf}.
For each distribution, events from simulated signal MC samples are used to generate a response matrix that accounts for bin-to-bin migration between the reconstruction-level and particle-level distributions. 

The statistical uncertainties of the unfolded distributions are estimated using pseudo-experiments, generated by fluctuating each bin of the observed spectrum according to a Poisson distribution with a mean value equal to the observed yield.
The shape uncertainties arising from the limited size of the signal MC sample are also obtained by generating pseudo-experiments.
The sources of systematic uncertainty are discussed in Section~\ref{xsec_intro}, with their impact on the unfolded distribution assessed by varying the response matrix
for each of the systematic uncertainty sources by one standard deviation and combining the resulting differences from the nominal values in quadrature.

\begin{figure}[hbtp]
\begin{center}
\includegraphics[width=0.45\textwidth]{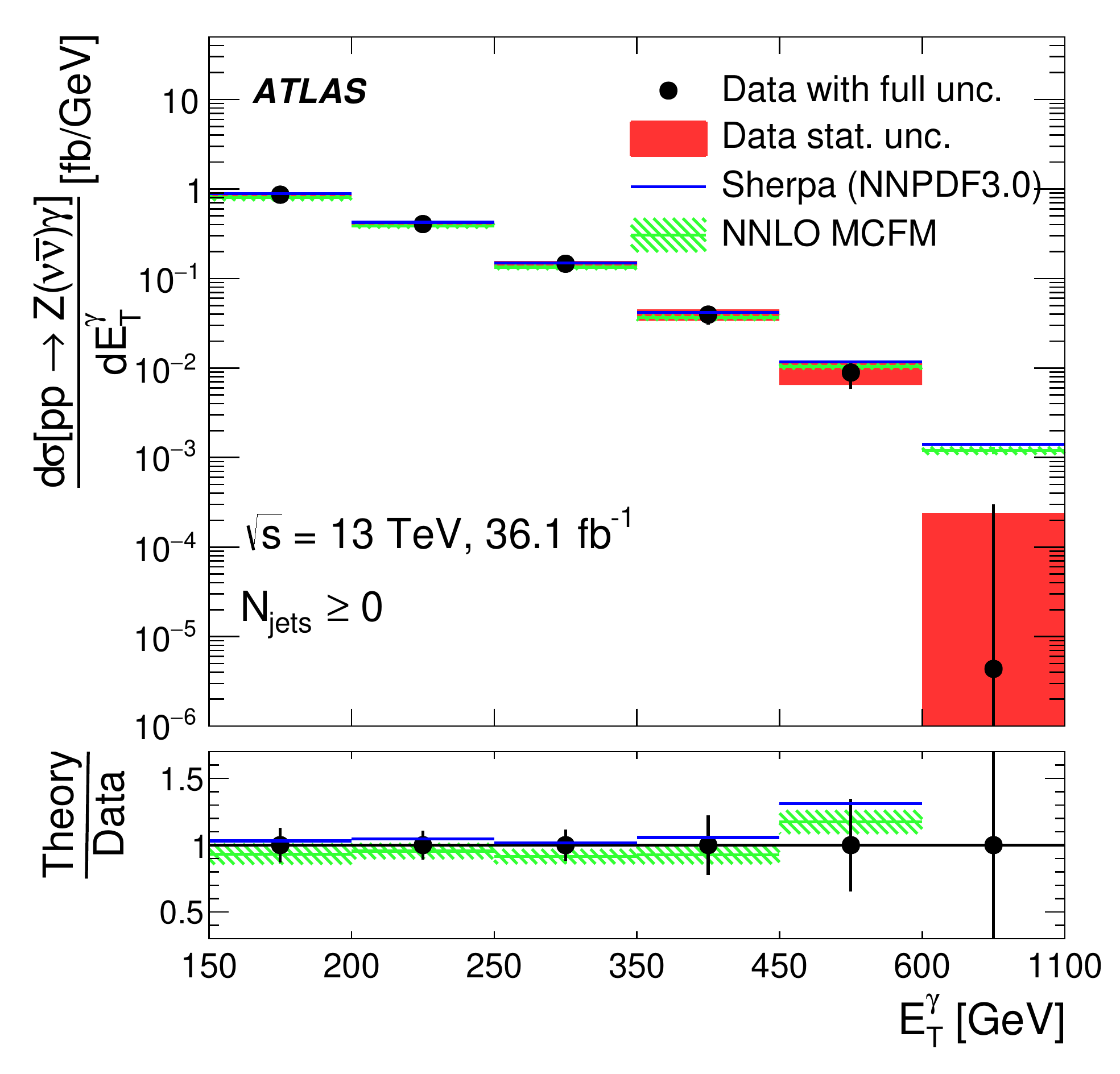}
\includegraphics[width=0.45\textwidth]{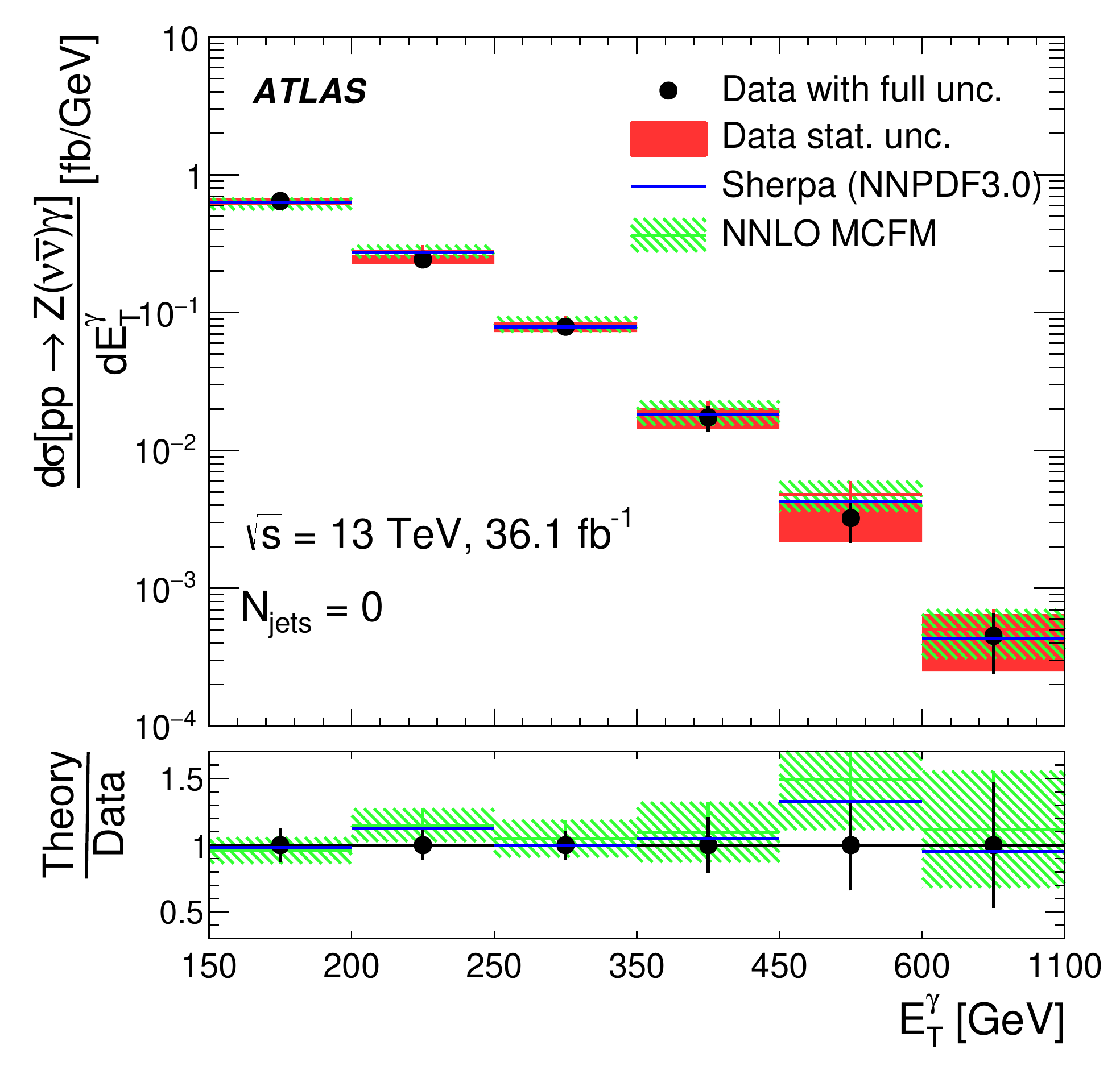}
\caption{The measured (points with error bars) and predicted differential cross sections as a function of $E_{\mathrm{T}}^{\gamma}$ for the  $pp \rightarrow Z(\nu\bar{\nu})\gamma$ process in the inclusive $N_{\mathrm{jets}} \geq 0$ (left) and exclusive $N_{\mathrm{jets}} = 0$ (right) extended fiducial regions.
The error bars on the data points show the sum in quadrature of the statistical and systematic uncertainties.
The \textsc{Mcfm} NNLO predictions are shown with shaded bands that indicate the theoretical uncertainties described in Section~\ref{sec:SMpredictions}.
For the \SHERPA predictions, systematic uncertainty is not considered, and the statistical uncertainties arising from the size of the MC samples are too small to be visible.
The lower plots show the ratios of the SM expectation to the measured values (shaded bands), with the error bars on the points showing the relative uncertainties in the experimental measurements. The bin size varies from 50~\GeV\ to 500~\GeV.}
\label{fig:UnfoldedPhotonEt}
\end{center}
\end{figure}

\begin{figure}[hbtp]
\begin{center}
\includegraphics[width=0.45\textwidth]{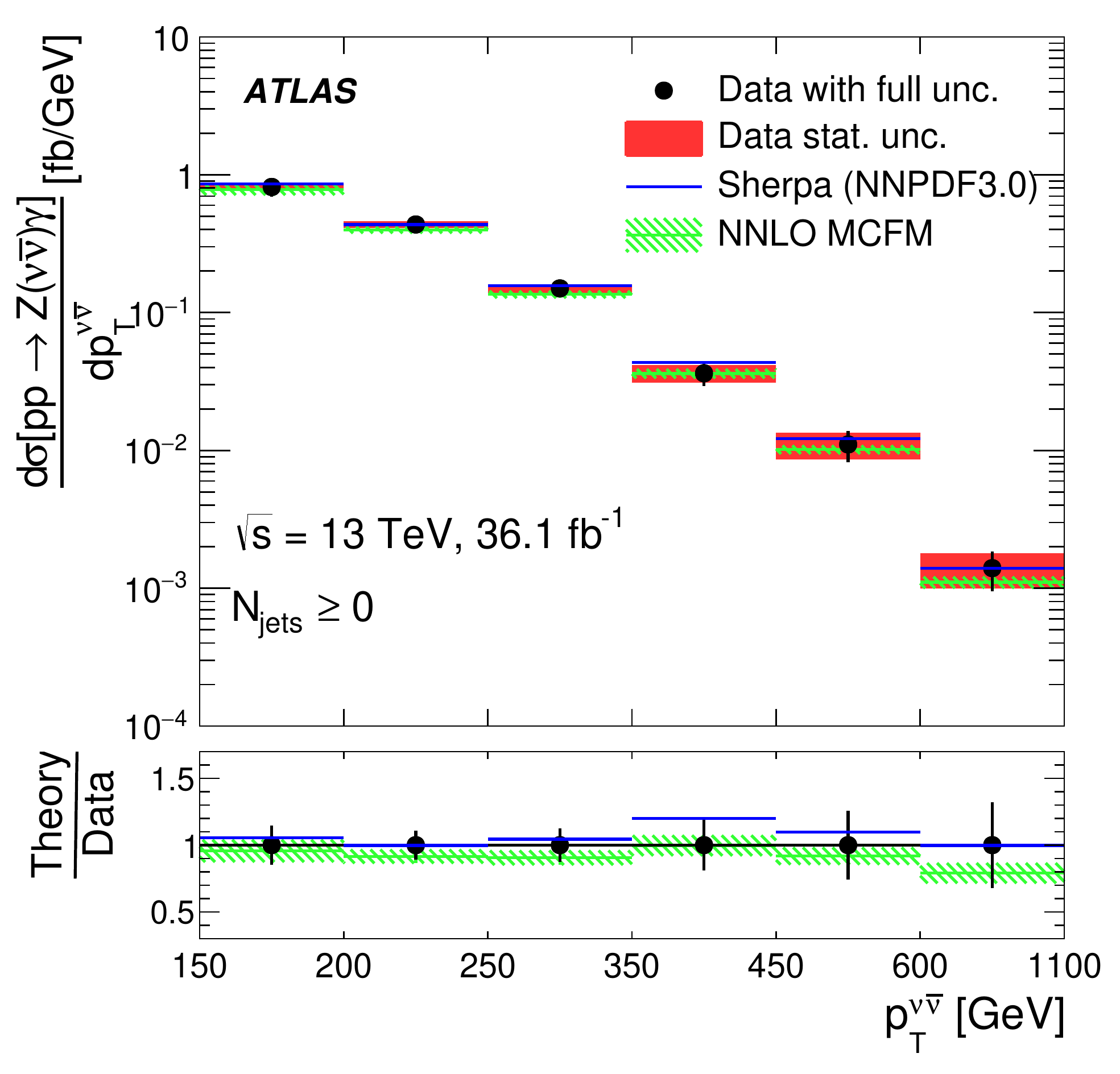}
\includegraphics[width=0.45\textwidth]{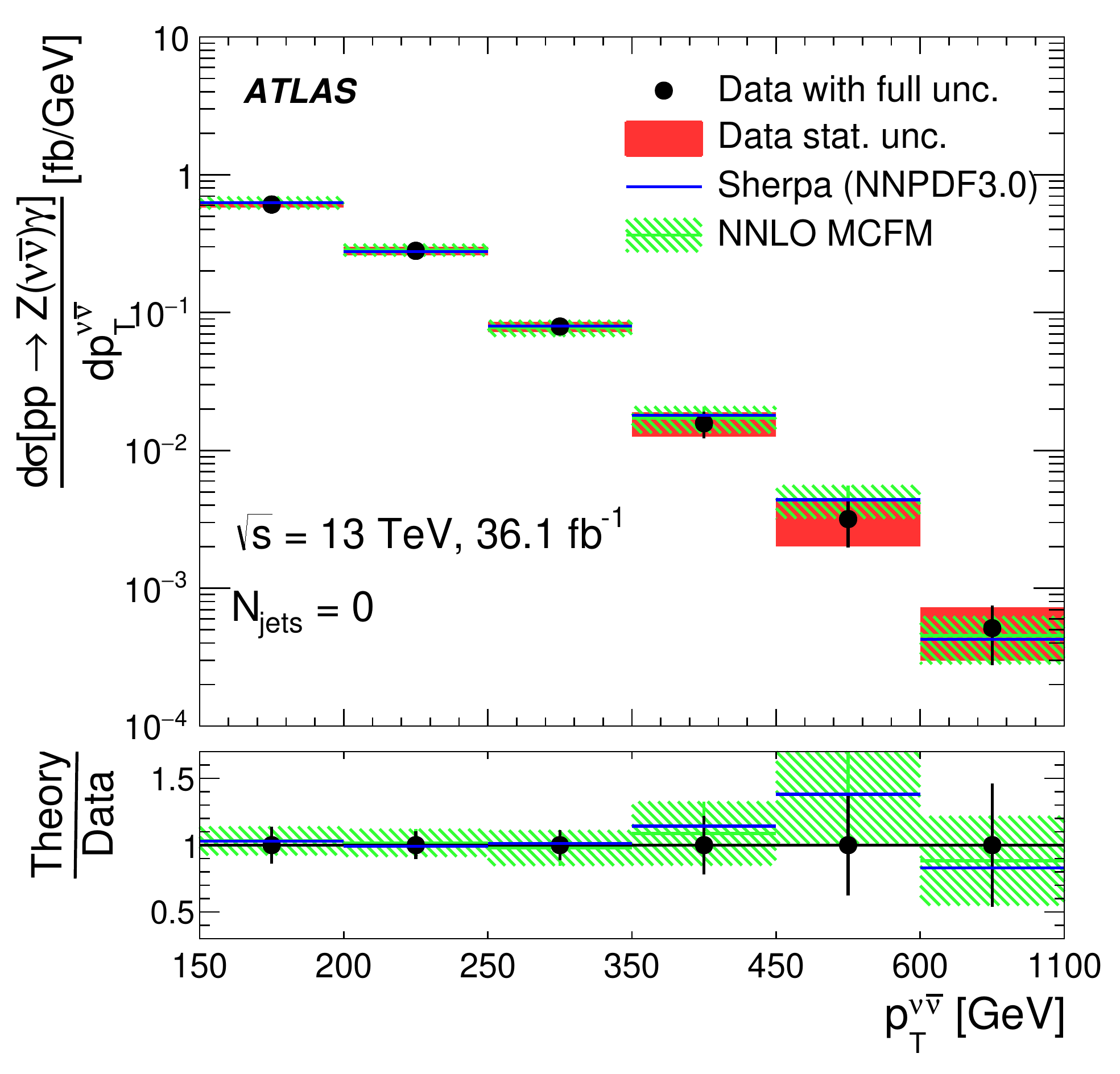}
\caption{The measured (points with error bars) and predicted differential cross sections as a function of $p_{\mathrm{T}}^{\nu\bar{\nu}}$ for the  $pp \rightarrow Z(\nu\bar{\nu})\gamma$ process in the inclusive $N_{\mathrm{jets}} \geq 0$ (left) and exclusive $N_{\mathrm{jets}} = 0$ (right) extended fiducial regions.
The error bars on the data points show the sum in quadrature of the statistical and systematic uncertainties.
The \textsc{Mcfm} NNLO predictions are shown with shaded bands that indicate the theoretical uncertainties described in Section~\ref{sec:SMpredictions}.
For the \SHERPA predictions, systematic uncertainty is not considered, and the statistical uncertainties arising from the size of the MC samples are too small to be visible.
The lower plots show the ratios of the SM expectation to the measured values (shaded bands), with the error bars on the points showing the relative uncertainties in the experimental measurements. The bin size varies from 50~\GeV\ to 500~\GeV.}
\label{fig:UnfoldedMET}
\end{center}
\end{figure}

\begin{figure}[hbtp]
\begin{center}
\includegraphics[width=0.45\textwidth]{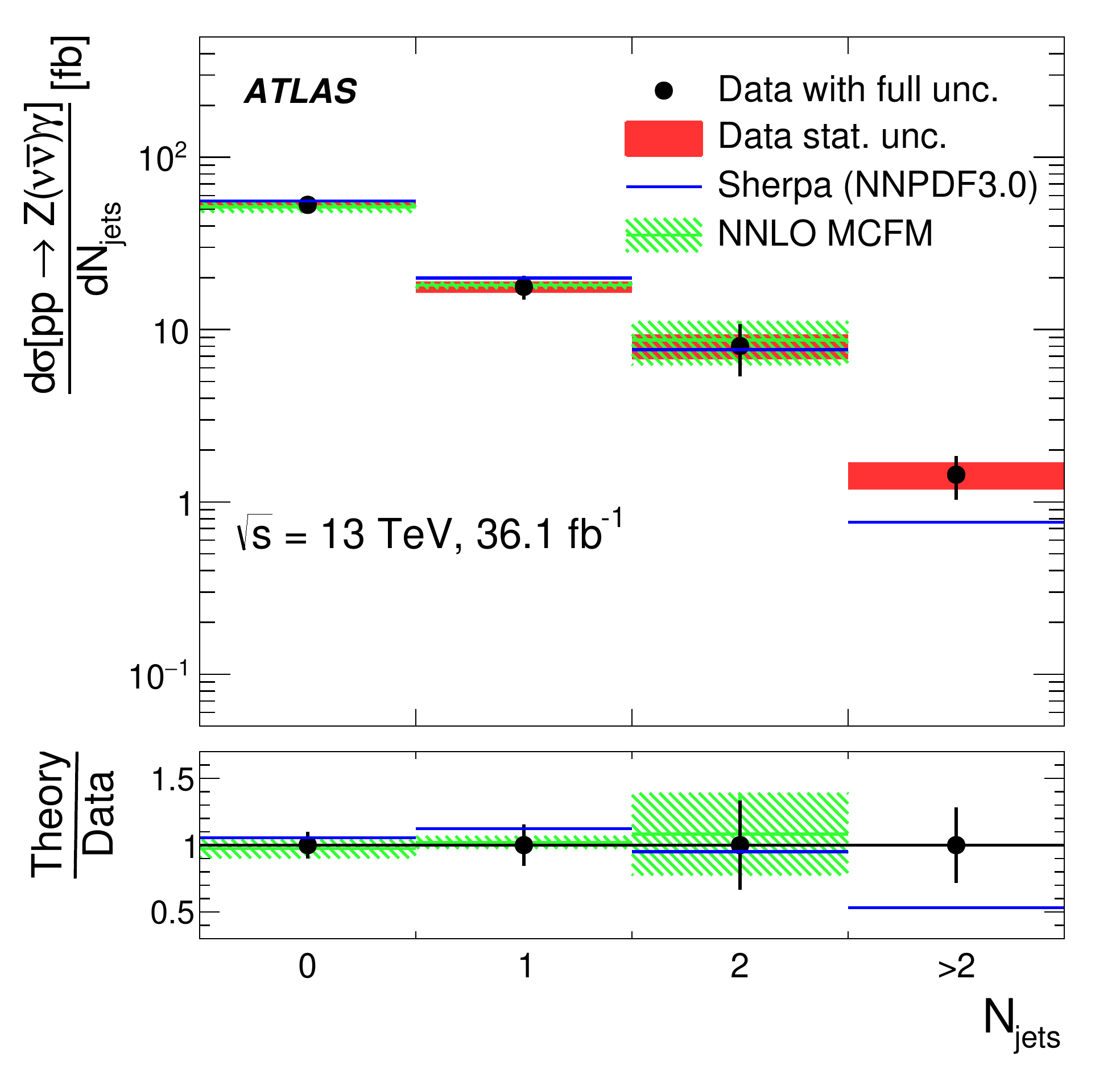}
\caption{ The measured (points with error bars) and predicted cross sections as a function of $N_{\mathrm{jets}}$ for the  $pp \rightarrow Z(\nu\bar{\nu})\gamma$ process in the extended fiducial region. The error bars on the data points show the sum in quadrature of the statistical and systematic uncertainties.
The \textsc{Mcfm} NNLO predictions are shown with shaded bands that indicate the theoretical uncertainties described in Section~\ref{sec:SMpredictions}.
For the \SHERPA predictions, systematic uncertainty is not considered, and the statistical uncertainties arising from the size of the MC samples are too small to be visible.
The lower plots show the ratios of the SM expectation to the measured values (shaded bands), with the error bars on the points showing the relative uncertainties in the experimental measurements.
}
\label{fig:UnfoldedNJets}
\end{center}
\end{figure}

The differential cross sections as a function of $E_{\mathrm{T}}^{\gamma}$ and $p_{\mathrm{T}}^{\nu\bar{\nu}}$ are shown in
Figures~\ref{fig:UnfoldedPhotonEt} and~\ref{fig:UnfoldedMET}, respectively, for both the inclusive and exclusive measurements.
Figure~\ref{fig:UnfoldedNJets} shows the cross section measured in bins of jet multiplicity.
The values of the SM expectations shown in the figures are obtained as described in Section~\ref{sec:SMpredictions}.

Good agreement with SM expectations is observed in all but the last bin of the $\ET^{\gamma}$ inclusive distribution. 
This disagreement is a consequence of the corresponding disagreement observed in Figure~\ref{fig:SR_inclusive}, which was investigated and found to be 
consistent with having arisen from a statistical fluctuation of the data.

%% file: atgc.tex
Vector-boson couplings are completely fixed within the Standard Model by the SU(2)$_{\mathrm{L}} \times $U(1)$_{\mathrm{Y}}$ gauge structure. Their measurement is thus a crucial test of the model. Any deviation from the SM prediction is referred to as an anomalous coupling.

Within the framework of the effective vertex function approach~\cite{PhysRevD.47.4889}, anomalous triple gauge-boson coupling contributions to $\Zboson\gamma$ production can be parameterized by four CP-violating ($h_1^V$, $h_2^V$) and four CP-conserving ($h_3^V$, $h_4^V$) complex parameters. Here the $V$ indices are $Z$ and $\gamma$, and $h_i^Z$ and $h_i^\gamma$ are the parameters of $ZZ\gamma$ and the $Z\gamma\gamma$ vertices, respectively.
The $h_{3}^{V}(h_{1}^{V})$ and $h_{4}^{V}(h_{2}^{V})$ parameters correspond to the electric (magnetic) dipole and magnetic (electric) quadrupole transition moments of $V$, respectively~\cite{ZPC_Zb_prop}.

All of these parameters are zero at tree level in the SM. Since the CP-conserving couplings $h_{3,4}^V$ do not interfere with the 
CP-violating couplings $h_{1,2}^V$, and their sensitivities to aTGCs are nearly identical~\cite{PhysRevD.47.4889}, the limits from this study are expressed solely in terms of the CP-conserving parameters $h_{3,4}^V$.

The yields of $\Zboson\gamma$ events with high $\ET^{\gamma}$ from the exclusive (zero-jet) selection are used to set limits on $h_{3,4}^V$.
The exclusive selection is used because it significantly reduces the SM contribution at high $\ET^{\gamma}$ and therefore optimizes the sensitivity to anomalous couplings.
The contribution from aTGCs increases with 
the $\ET$ of the photon, and the measurement of $\Zboson\gamma$ production is found to have the highest sensitivity to aTGCs by restricting the search to the portion of the extended fiducial region with $\ET^{\gamma}$ greater than 600~\GeV.

Cross-section values modified by the inclusion of aTGCs ($\sigma_{\Zboson\gamma}^{\mathrm{aTGC}}$) are obtained from the \textsc{Mcfm} generator. 
These values are displayed in Table~\ref{table:nunugXsecATGC} for several combinations of choices of the $ZZ\gamma$ vertex parameters
$h_3^Z$ and $h_4^Z$. 

\begin{table}
\begin{center}
\begin{tabular}{r@{.}l| r@{.}lr@{.}lr@{.}l}
\hline
\multicolumn{2}{l|}{\diagbox[width=5em]{$h_3^Z$}{$h_4^Z$}} & \multicolumn{2}{l}{$-5\cdot10^{-7}$}  & \multicolumn{2}{l}{0}  & \multicolumn{2}{l}{$5\cdot10^{-7}$} \\
\hline
\multicolumn{2}{l|}{$-5\cdot10^{-4}$} & $\ \:\,$0&439  & 0&696  & 1&42 \\
\multicolumn{2}{l|}{$\ \:\,0$} & $\ \:\,$0&477  & 0&243  & 0&483 \\
\multicolumn{2}{l|}{$\ \:\,5\cdot10^{-4}$}  & $\ \:\,$1&40  & 0&674  & 0&424 \\
\hline
\end{tabular}
\end{center}
\caption{Cross sections [fb] for the exclusive $Z(\nu\bar{\nu})\gamma$ process, requiring a photon with $\ET^{\gamma}>600$~\GeV, for different values of $h_3^Z$ (vertical), and $h_4^Z$ (horizontal). For the Standard Model with no anomalous triple gauge-boson couplings, $h_3^Z = h_4^Z = 0$.}
\label{table:nunugXsecATGC}
\end{table}

The expected number of $\Zboson\gamma$ events in the aTGC region ($N^{\mathrm{aTGC}}_{\Zboson\gamma}(h_3^V,h_4^V)$, where $V = Z$ or $\gamma$) is obtained using

\begin{equation}
N^{\mathrm{aTGC}}_{\Zboson\gamma}(h_3^V,h_4^V) = \sigma_{\Zboson\gamma}^{\mathrm{aTGC}}(h_3^V,h_4^V) \cdot C_{\Zboson\gamma} \cdot A_{\Zboson\gamma} \cdot C^{*(\mathrm{parton}\to\mathrm{particle})} \cdot \int L\,\mathrm{d}t.
\label{eqn:NaTGC}
\end{equation}

The anomalous couplings influence the kinematic properties of the $\Zboson\gamma$ events and thus the efficiency factor of the event reconstruction ($C_{\Zboson
\gamma}$). The maximum variation of $C_{\Zboson\gamma}$ due to non-zero aTGC parameters within the aTGC limits measured in this paper (about 7\%) is adopted as an additional systematic uncertainty. The effect of anomalous couplings on the acceptance factor ($A_{\Zboson\gamma}$) and parton-to-particle factor ($C^{*(\mathrm{parton}\to \mathrm{particle})}$) is an order of magnitude smaller than that on $C_{\Zboson\gamma}$, and so is neglected.

Limits on a given aTGC parameters are extracted from a frequentist profile-likelihood test similar to that of Section~\ref{xsec_pheno}. The profile likelihood depends on the observed number of exclusive $\Zboson\gamma$ candidate events, the amount of expected signal as a function of aTGC given by Eq.~(\ref{eqn:NaTGC}), and the estimated number of background events. 
A point in the aTGC space is accepted (rejected) at the 95\% confidence level (CL) if fewer (more) than 95\% of randomly generated pseudo-experiments exhibit larger profile-likelihood ratio values than that observed in data. In this context, a pseudo-experiment is a set of randomly generated numbers of events that follow a Poisson distribution with mean equal to the sum of the number of expected signal events and the estimated number of background events. Systematic uncertainties are incorporated into the pseudo-experiments via a set of nuisance parameters with correlated Gaussian constraints. All nuisance parameters are allowed to fluctuate in the pseudo-experiments.

No evidence of anomalous couplings is observed. The allowed 95\% CL ranges for the anomalous couplings are shown in Table~\ref{table:observedExpected1DLimitsATGC} for $\Zboson\Zboson\gamma$ ($h_3^Z$ and $h_4^Z$) and the $\Zboson\gamma\gamma$ ($h_3^\gamma$ and $h_4^\gamma$) vertices. 
Limits on anomalous couplings imposed by this analysis are $3$--$7$ times more stringent
than those from prior studies~\cite{STDM-2014-01}.

\begin{table}
\begin{center}
\begin{tabular}{lll}
\hline
Parameter & \multicolumn{2}{c}{Limit 95\% CL}\\
 & \multicolumn{1}{c}{Measured} & \multicolumn{1}{c}{Expected} \\
 \hline
$h_3^\gamma$ &  $(-3.7 \times 10^{-4}, 3.7 \times 10^{-4})$ & $(-4.2 \times 10^{-4}, 4.3 \times 10^{-4}) $\\
$h_3^Z$ & $(-3.2 \times 10^{-4}, 3.3 \times 10^{-4})$  & $(-3.8 \times 10^{-4}, 3.8 \times 10^{-4}) $\\
$h_4^\gamma$ &  $(-4.4 \times 10^{-7}, 4.3 \times 10^{-7})$ &  $(-5.1 \times 10^{-7}, 5.0 \times 10^{-7}) $\\
$h_4^Z$ & $(-4.5 \times 10^{-7}, 4.4 \times 10^{-7})$ & $(-5.3 \times 10^{-7}, 5.1 \times 10^{-7}) $\\

\hline
\end{tabular}
\end{center}
\caption{Observed and expected one-dimensional 95\% CL limits on $h_3^{\gamma}$, $h_3^Z$, $h_4^{\gamma}$ and $h_4^Z$,
assuming that any observed excess in data relative to the associated SM estimate 
is due solely to $h_3^V$ or $h_4^V$. For each row, all parameters
other than the one under study are set to 0.}
\label{table:observedExpected1DLimitsATGC}
\end{table}

\begin{figure}
\begin{center}
\includegraphics[width=0.45\textwidth]{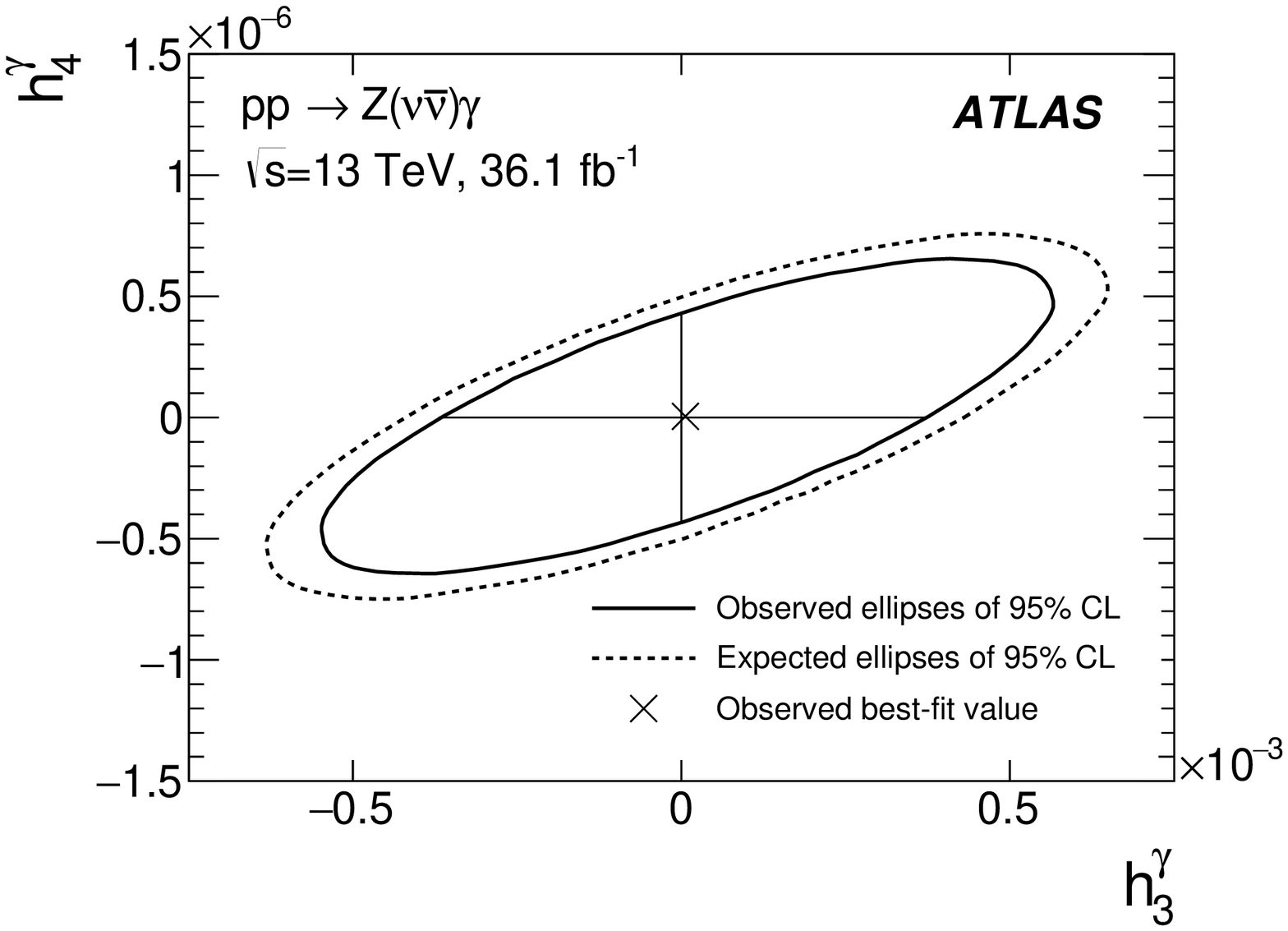}
\includegraphics[width=0.45\textwidth]{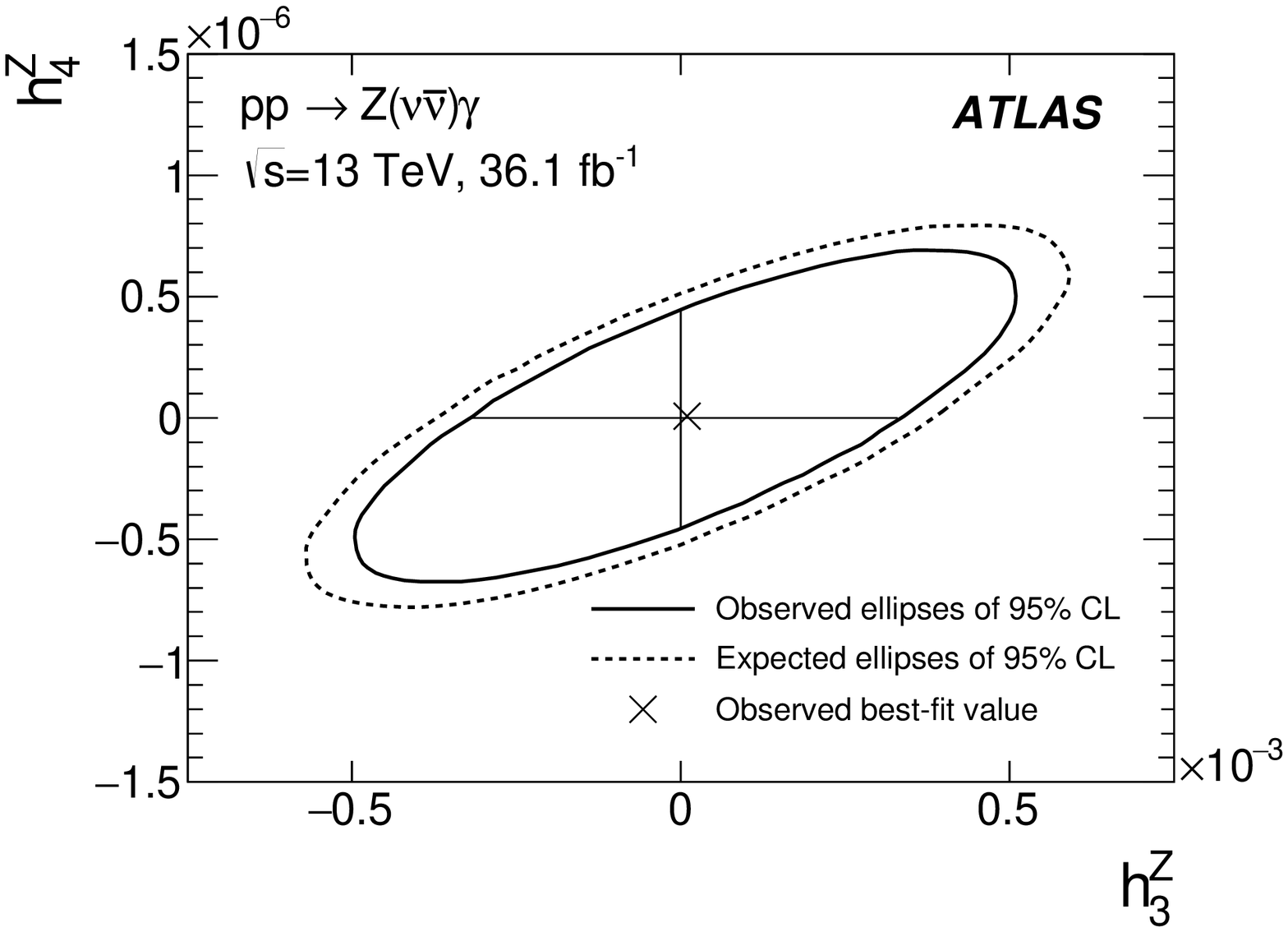}
\caption{Observed (solid) and expected (dashed) ellipses of 95\% CL on the linear combinations of the pairs of anomalous couplings $h_3^{\gamma}$ and $h_4^{\gamma}$ (left) and $h_3^Z$ and $h_4^Z$ (right). The horizontal and vertical lines inside each contour correspond to the limits found in the one-parameter fit procedure, while the orientation of the ellipses indicates the correlations between the parameters in the two-dimensional fit. In each case, the two parameters not being displayed are set to 0.}
\label{fig:TwoDimensionalLimits_inf}
\end{center}
\end{figure}

Limits on possible combinations of each pair of aTGC parameters are also evaluated. The ellipses of 95\% CL on linear combination of the pairs of anomalous couplings are shown on the $(h_3^{\gamma}, h_4^{\gamma})$ and $(h_3^{\Zboson}, h_4^{\Zboson})$ planes in Figure~\ref{fig:TwoDimensionalLimits_inf}, which are the only such pairs that are expected to interfere~\cite{PhysRevD.47.4889}.

Allowed ranges are also determined for parameters of the effective field theory (EFT) of Ref.~\cite{bib:eft}, which includes four dimension-8 operators describing aTGC interactions of neutral gauge bosons. The coefficients of these operators are denoted $C_{\widetilde{B}W}/\Lambda^4$, $C_{BW}/\Lambda^4$, $C_{WW}/\Lambda^4$ and $C_{BB}/\Lambda^4$, as described in Ref.~\cite{bib:eft_nTGC}. The parameter $\Lambda$ has the dimension of mass and is associated with the energy scale of the new physics described by the EFT.
The 95\% CL limits on these EFT parameters displayed in Table~\ref{table:observedExpected1DLimitsATGC_EFT} are derived from the limits of
Table~\ref{table:observedExpected1DLimitsATGC} making use of a linear transformation relating the EFT and vertex function aTGC parameters, obtained from Ref.~\cite{bib:eft_nTGC}.

\begin{table}
\begin{center}
\begin{tabular}{l r@{, }lr@{, }l}
\hline
Parameter & \multicolumn{4}{c}{Limit 95\% CL}\\
  &  \multicolumn{2}{c}{\hspace{1mm}Measured [\TeV$^{-4}$]}&  \multicolumn{2}{c}{\hspace{2mm}Expected [\TeV$^{-4}$]} \\ \hline
$C_{\widetilde{B}W}/\Lambda^4$ &\hspace{3mm}  $(-1.1$ & $1.1 )$ & $(-1.3$& $1.3) $\\
$C_{BW}/\Lambda^4$ &\hspace{3mm} $(-0.65$& $0.64)$  &\hspace{3mm} $(-0.74$& $0.74) $\\
$C_{WW}/\Lambda^4$ &\hspace{3mm}  $(-2.3$& $2.3)$ &\hspace{3mm}  $(-2.7$& $2.7) $\\
$C_{BB}/\Lambda^4$ &\hspace{3mm} $(-0.24$&$0.24)$ &\hspace{3mm} $(-0.28$&$0.27) $\\

\hline
\end{tabular}
\end{center}
\caption{Observed and expected one-dimensional 95\% CL limits on the $C_{\widetilde{B}W}/\Lambda^4$, $C_{BW}/\Lambda^4$, $C_{WW}/\Lambda^4$ and $C_{BB}/\Lambda^4$ EFT parameters, assuming that any excess in data over the SM expectation is due solely to a non-zero value of the parameter $C_{\widetilde{B}W}/\Lambda^4$, $C_{BW}/\Lambda^4$, $C_{WW}/\Lambda^4$ or $C_{BB}/\Lambda^4$. For each row, all parameters other than the one under study are set to 0.}
\label{table:observedExpected1DLimitsATGC_EFT}
\end{table}

%% file: summary.tex
The cross section for the production of a $Z$ boson in association with an isolated high-energy photon is measured using 36.1~$\ifb$ of $pp$ collisions at $\sqrt{s} = 13$~\TeV\ collected 
with the ATLAS detector at the LHC. The analysis uses the invisible decay mode $\Zboson\to\nnbar$ of the $Z$ boson, and is performed in a fiducial phase space closely matching the detector acceptance.

Kinematic distributions are presented in terms of differential cross sections as a function of
the transverse energy of the photon, the missing transverse momentum, and the jet multiplicity.
Measurements are made for both the inclusive case, with no requirements on the system recoiling against the $\Zboson\gamma$ pair, and the exclusive case 
in which no jets with $\pT > 50$~\GeV\ are allowed within $|\eta|< 4.5$.

The results are compared with SM expectations derived from a parton shower Monte Carlo generator ($\SHERPA$) and from parton-level perturbative calculations carried out at NNLO ({\textsc{Mcfm}} and {\textsc{Matrix}}). Good agreement is observed between the measured and expected total and differential cross sections. 

In the absence of significant deviations from SM expectations, the data are used to set limits on anomalous couplings of photons and $\Zboson$ bosons. 
Limits on aTGCs are determined using a modified SM Lagrangian that includes operators proportional to the $h_3^V$ and $h_4^V$ ($V = Z$ or $\gamma$) parameters of the vertex function parameterization of aTGC contributions to $\Zboson\gamma$ production. The limits are also transformed into limits on the $C_{\widetilde{B}W}/\Lambda^4$, $C_{BW}/\Lambda^4$, $C_{WW}/\Lambda^4$ and $C_{BB}/\Lambda^4$ parameters of an effective field theory formulation of aTGC effects. The limits obtained from the current study are $3$--$7$ times more stringent than those available prior to this study.

%% file: acknowledgements/Acknowledgements.tex
% Acknowledgements for papers with collision data
% Version 05-Oct-2018

% Standard acknowledgements start here
%----------------------------------------------
We thank CERN for the very successful operation of the LHC, as well as the
support staff from our institutions without whom ATLAS could not be
operated efficiently.

We acknowledge the support of ANPCyT, Argentina; YerPhI, Armenia; ARC, Australia; BMWFW and FWF, Austria; ANAS, Azerbaijan; SSTC, Belarus; CNPq and FAPESP, Brazil; NSERC, NRC and CFI, Canada; CERN; CONICYT, Chile; CAS, MOST and NSFC, China; COLCIENCIAS, Colombia; MSMT CR, MPO CR and VSC CR, Czech Republic; DNRF and DNSRC, Denmark; IN2P3-CNRS, CEA-DRF/IRFU, France; SRNSFG, Georgia; BMBF, HGF, and MPG, Germany; GSRT, Greece; RGC, Hong Kong SAR, China; ISF and Benoziyo Center, Israel; INFN, Italy; MEXT and JSPS, Japan; CNRST, Morocco; NWO, Netherlands; RCN, Norway; MNiSW and NCN, Poland; FCT, Portugal; MNE/IFA, Romania; MES of Russia and NRC KI, Russian Federation; JINR; MESTD, Serbia; MSSR, Slovakia; ARRS and MIZ\v{S}, Slovenia; DST/NRF, South Africa; MINECO, Spain; SRC and Wallenberg Foundation, Sweden; SERI, SNSF and Cantons of Bern and Geneva, Switzerland; MOST, Taiwan; TAEK, Turkey; STFC, United Kingdom; DOE and NSF, United States of America. In addition, individual groups and members have received support from BCKDF, the Canada Council, CANARIE, CRC, Compute Canada, FQRNT, and the Ontario Innovation Trust, Canada; EPLANET, ERC, ERDF, FP7, Horizon 2020 and Marie Sk{\l}odowska-Curie Actions, European Union; Investissements d'Avenir Labex and Idex, ANR, R{\'e}gion Auvergne and Fondation Partager le Savoir, France; DFG and AvH Foundation, Germany; Herakleitos, Thales and Aristeia programmes co-financed by EU-ESF and the Greek NSRF; BSF, GIF and Minerva, Israel; BRF, Norway; CERCA Programme Generalitat de Catalunya, Generalitat Valenciana, Spain; the Royal Society and Leverhulme Trust, United Kingdom.

The crucial computing support from all WLCG partners is acknowledged gratefully, in particular from CERN, the ATLAS Tier-1 facilities at TRIUMF (Canada), NDGF (Denmark, Norway, Sweden), CC-IN2P3 (France), KIT/GridKA (Germany), INFN-CNAF (Italy), NL-T1 (Netherlands), PIC (Spain), ASGC (Taiwan), RAL (UK) and BNL (USA), the Tier-2 facilities worldwide and large non-WLCG resource providers. Major contributors of computing resources are listed in Ref.~\cite{ATL-GEN-PUB-2016-002}.
%----------------------------------------------

%% file: atlas_authlist.tex
% ATLAS Collaboration author list
% Reference date of STDM-2017-18 is 2018-06-17
% Author list last updated on date 08-AUG-18
% Data extracted on 08-Aug-2018 for paper reference STDM-2017-18
% at 5:56pm
 
\begin{flushleft}
{\Large The ATLAS Collaboration}

\bigskip

M.~Aaboud$^\textrm{\scriptsize 34d}$,    
G.~Aad$^\textrm{\scriptsize 99}$,    
B.~Abbott$^\textrm{\scriptsize 124}$,    
O.~Abdinov$^\textrm{\scriptsize 13,*}$,    
B.~Abeloos$^\textrm{\scriptsize 128}$,    
D.K.~Abhayasinghe$^\textrm{\scriptsize 91}$,    
S.H.~Abidi$^\textrm{\scriptsize 164}$,    
O.S.~AbouZeid$^\textrm{\scriptsize 39}$,    
N.L.~Abraham$^\textrm{\scriptsize 153}$,    
H.~Abramowicz$^\textrm{\scriptsize 158}$
% ...
\bigskip
\\

% ...
$^{13}$Institute of Physics, Azerbaijan Academy of Sciences, Baku; Azerbaijan.\\
% ...
$^{34}$$^{(a)}$Facult\'e des Sciences Ain Chock, R\'eseau Universitaire de Physique des Hautes Energies - Universit\'e Hassan II, Casablanca;$^{(b)}$Centre National de l'Energie des Sciences Techniques Nucleaires (CNESTEN), Rabat;$^{(c)}$Facult\'e des Sciences Semlalia, Universit\'e Cadi Ayyad, LPHEA-Marrakech;$^{(d)}$Facult\'e des Sciences, Universit\'e Mohamed Premier and LPTPM, Oujda;$^{(e)}$Facult\'e des sciences, Universit\'e Mohammed V, Rabat; Morocco.\\
% ...
$^{39}$Niels Bohr Institute, University of Copenhagen, Copenhagen; Denmark.\\
% ...
$^{91}$Department of Physics, Royal Holloway University of London, Egham; United Kingdom.\\
% ...
$^{99}$CPPM, Aix-Marseille Universit\'e, CNRS/IN2P3, Marseille; France.\\
% ...
$^{124}$Homer L. Dodge Department of Physics and Astronomy, University of Oklahoma, Norman OK; United States of America.\\
% ...
$^{128}$LAL, Universit\'e Paris-Sud, CNRS/IN2P3, Universit\'e Paris-Saclay, Orsay; France.\\
% ...
$^{153}$Department of Physics and Astronomy, University of Sussex, Brighton; United Kingdom.\\
% ...
$^{158}$Raymond and Beverly Sackler School of Physics and Astronomy, Tel Aviv University, Tel Aviv; Israel.\\
% ...
$^{164}$Department of Physics, University of Toronto, Toronto ON; Canada.\\
% ...
$^{a}$ Also at Borough of Manhattan Community College, City University of New York, NY; United States of America.\\
$^{b}$ Also at Centre for High Performance Computing, CSIR Campus, Rosebank, Cape Town; South Africa.\\
$^{c}$ Also at CERN, Geneva; Switzerland.\\
$^{d}$ Also at CPPM, Aix-Marseille Universit\'e, CNRS/IN2P3, Marseille; France.\\
$^{e}$ Also at D\'epartement de Physique Nucl\'eaire et Corpusculaire, Universit\'e de Gen\`eve, Gen\`eve; Switzerland.\\
% ...
$^{*}$ Deceased

\end{flushleft}

% Created with Glance <Atlas.Glance@cern.ch>